\documentclass[lettersize,journal]{IEEEtran}
\usepackage{amsmath,amssymb,amsfonts}
\usepackage{textcomp}
\usepackage{algpseudocode}
\usepackage[linesnumbered,ruled,vlined]{algorithm2e}
\usepackage{amsthm}
\usepackage{array}
\usepackage{color}
\usepackage{cases}
\usepackage{epsfig}
\usepackage{extarrows}
\usepackage{graphicx} 
\usepackage{graphics}
\usepackage{stfloats}
\usepackage{subfig}
\usepackage{float}
\usepackage{setspace}
\usepackage{bm}
\usepackage{xparse}
\usepackage{cite}
\usepackage{amsmath,amsfonts,amssymb,mathrsfs}
\usepackage{nccmath}
\usepackage{booktabs}
\usepackage[font={small}]{caption}
\usepackage{mdframed}
\usepackage{catchfilebetweentags}
\usepackage{hyperref}
\hypersetup{hidelinks,
	colorlinks=true,
	allcolors=black,
	pdfstartview=Fit,
	breaklinks=true
}
\usepackage[outdir=./fig/epstopdf_converted/]{epstopdf}
\usepackage{multirow}
\usepackage{diagbox}
\usepackage{makecell}
\usepackage{verbatim}
\usepackage{subcaption}

\newcommand{\tabincell}[2]{\begin{tabular}{@{}#1@{}}#2\end{tabular}}

\def\sfT{\mathsf{T}}
\def\sfH{\mathsf{H}}
\def\argmin{\mathop{\arg\min}}
\def\argmax{\mathop{\arg\max}}

\def\rmref{\mathrm{ref}}
\def\rmsur{\mathrm{sur}}
\def\rmtar{\mathrm{tar}}

\def\rmbist{\mathrm{bist}}
\def\rmtx{\mathrm{t}}
\def\rmrx{\mathrm{r}}
\def\rmtxrx{\rmtx\textrm{--}\rmrx}
\def\rmD{\mathrm{D}}
\def\rmcf{\mathrm{cf}}
\def\rmtotal{\mathrm{total}}
\def\rmguard{\mathrm{guard}}
\def\rmtrain{\mathrm{train}}
\def\rmtt{\mathrm{tt}}
\def\rmct{\mathrm{ct}}
\def\rminit{\mathrm{init}}
\def\rmcnf{\mathrm{cnf}}
\def\rmdel{\mathrm{del}}
\def\rmrem{\mathrm{rem}}
\def\rmMah{\mathrm{Mah}}

\DeclareMathOperator{\atantwo}{arctan2}
\DeclareMathOperator{\diag}{diag}

\IEEEoverridecommandlockouts

\begin{document}

\title{
	An Experimental Study of Passive UAV Tracking\\
	with Digital Arrays and Cellular Downlink Signals
}

\author{
	Yifei~Sun\IEEEauthorrefmark{1,2},
	Chao~Yu\IEEEauthorrefmark{1},
	Yan~Luo\IEEEauthorrefmark{1},
	Tony~Xiao~Han\IEEEauthorrefmark{3},
	Haisheng~Tan\IEEEauthorrefmark{4},
	Rui~Wang\IEEEauthorrefmark{1},
	Francis~C.~M.~Lau\IEEEauthorrefmark{2}
	\thanks{
		Yifei Sun is with the Department of Electrical and Electronic Engineering, Southern University of Science and Technology, Shenzhen 518055, China, and also the Department of Computer Science, School of Computing and Data Science, The University of Hong Kong, Hong Kong, China.
		Chao Yu, Yan Luo and Rui Wang are with the Department of Electrical and Electronic Engineering, Southern University of Science and Technology, Shenzhen 518055, China.
		Tony Xiao Han is with Huawei Technologies Company Ltd., Shenzhen 518063, China.
		Haisheng Tan is with School of Artificial Intelligence and Data Science, University of Science and Technology of China, Hefei 230027, China.
		Francis C. M. Lau is with the Department of Computer Science, School of Computing and Data Science, The University of Hong Kong, Hong Kong, China. (Corresponding author: Rui Wang.)
	}
}

\maketitle

\begin{abstract}
	Given the prospects of the low-altitude economy (LAE) and the popularity of unmanned aerial vehicles (UAVs), there are increasing demands on monitoring flying objects at low altitude in wide urban areas.
	In this work, the widely deployed long-term evolution (LTE) base station (BS) is exploited to illuminate UAVs in bistatic trajectory tracking.
	Specifically, a passive sensing receiver with two digital antenna arrays is proposed and developed to capture both the line-of-sight (LoS) signal and the scattered signal off a target UAV.
	From their cross ambiguity function, the bistatic range, Doppler shift and angle-of-arrival (AoA) of the target UAV can be detected in a sequence of time slots.
	In order to address missed detections and false alarms of passive sensing, a multi-target tracking framework is adopted to track the trajectory of the target UAV.
	It is demonstrated by experiments that the proposed UAV tracking system can achieve a meter-level accuracy.
\end{abstract}

\IEEEpeerreviewmaketitle


\section{Introduction}
\label{sec:introduction}
The emerging applications of unmanned aerial vehicles (UAVs) have boosted the prosperity of the low-altitude economy (LAE).
However, the popularity of UAVs can also pose security risks, especially due to illegal or unauthorized flights in urban areas.
Hence, there are increasing demands on the detection, localization and tracking of non-cooperative UAVs.
Although active integrated sensing and communications (ISAC) technology that exploits the widely deployed communication infrastructure in sensing is a promising solution for monitoring flying objects at low altitudes \cite{9737357,cheng2024networkedisaclowaltitudeeconomy}, upgrading base stations (BS) to support joint communication and sensing faces significant challenges in interference management, resource allocation, and deployment cost \cite{khawaja2024surveydetectionclassificationtracking}.
Therefore, the passive-sensing solution that utilizes the existing cellular infrastructure as the illuminators becomes appealing.
Passive sensing with one illuminator and one receiver is referred to as bistatic sensing, and sensing with more than one illuminator or receiver is referred to as multi-static sensing.

There have been plenty of works focusing on the detection or tracking of airplanes or UAVs with passive sensing, which used various types of broadcast signals, including frequency modulation (FM) broadcasting \cite{8835773,:/content/journals/10.1049/ip-rsn_20045077}, digital audio broadcasting (DAB) \cite{7944357,4721007} and digital video broadcasting (DVB) \cite{4720984,8950339}, as the illuminator.
For example in \cite{8835773}, the FM broadcast signal was exploited in bistatic sensing to detect the bistatic range, bistatic Doppler shift and angle of arrival (AoA) of a commercial aircraft in an airport.
In \cite{8950339}, a passive multi-static radar was implemented to track a UAV by exploiting  multiple DVB signal emitters in a time-division-multi-frequency manner.
Moreover, the BS of the global system for mobile communications (GSM) was also exploited as the illuminator in passive sensing \cite{5307190,7411917,9266673,7497375}.
For example, a track-before-detect method with a particle filter was proposed to track a quadcopter by collecting echo signals from five GSM BSs in a multi-static sensing configuration.
In all the above works, the bandwidths of illuminating signals were narrow, leading to low-resolution range estimations.
Furthermore, due to the large wavelengths of some broadcast signals (e.g., 2.78 m for FM, 1.25 m for DAB, and 0.35 m for DVB-T), it was infeasible to employ a large-scale antenna array on a portable system with limited space.
Hence, the angular resolutions of the related sensing systems were low, which resulted in kilometer-level localization error in bistatic sensing \cite{8835773,:/content/journals/10.1049/ip-rsn_20045077}.
The only way to improve the trajectory tracking accuracy was to invest more receivers or more radio frequency (RF) chains at the receiver, such that the measurements at more AoAs could be simultaneously obtained via multi-static sensing \cite{8950339,9266673}.

To achieve better range resolution through larger bandwidth, the long-term evolution (LTE) and fifth-generation mobile communication systems (5G) were recently considered the illuminator \cite{https://doi.org/10.1049/joe.2019.0583,10.1145/3507657.3529658,9764210,rs14236146}.
For example, in \cite{https://doi.org/10.1049/joe.2019.0583} and \cite{rs14236146}, the LTE and 5G signals were exploited, respectively, to detect a UAV in a given direction, where a highly directional antenna was used to acquire a sufficient signal-to-clutter-plus-noise ratio (SCNR). However, this approach might miss the target if the UAV is out of the field of view (FoV) of the directional antenna.
One method to address the above issue is to densely deploy multiple illuminators and sensing receivers for multi-static sensing.
It was shown in \cite{10.1145/3507657.3529658} that a sub-meter localization error could be achieved with multi-static sensing and directional antennas.
Moreover, in \cite{9764210}, a sensor fusion technology integrating the multi-static sensing with the out-of-band information of electro-optical and near-infrared cameras was proposed.
However, it is still unknown whether bistatic sensing based on LTE or 5G signals could provide sufficient range and AoA resolutions for the trajectory tracking of a UAV.
\begin{table*}[tb]
	\centering
	\resizebox{\linewidth}{!}{%
		\begin{tabular}{|c|c|c|c|c|c|c|c|c|c|c|c|c|}
			\hline
			Ref.                                              & IO                                     & \tabincell{c}{Fre-                                                                                                                                                                                                                                                                                            \\quency \\(Hz)}                          & \tabincell{c}{Band-\\width \\(Hz)}
			                                                  & \tabincell{c}{Bistatic (B)                                                                                                                                                                                                                                                                                                                             \\/Multi-\\static (M)}
			                                                  & \tabincell{c}{Detection (D)                                                                                                                                                                                                                                                                                                                            \\/Localization (L)\\/Tracking (T)}
			                                                  & \tabincell{c}{Surveillance                                                                                                                                                                                                                                                                                                                             \\antenna}
			                                                  & \tabincell{c}{Mea-                                                                                                                                                                                                                                                                                                                                     \\rmsure-                                                                                                                                                                                                                                                                                                                                                         \\ments}
			                                                  & \tabincell{c}{Tracking                                                                                                                                                                                                                                                                                                                                 \\methods}
			                                                  & Target
			                                                  & \tabincell{c}{Baseline                                                                                                                                                                                                                                                                                                                                 \\length}
			                                                  & \tabincell{c}{Target                                                                                                                                                                                                                                                                                                                                   \\range}
			                                                  & \tabincell{c}{Locali-                                                                                                                                                                                                                                                                                                                                  \\zation                                                                                                                                                                                                                                                                                                                                                     \\error}\\
			\hline
			\multirow{2}{*}{\cite{8835773}}                   & \multirow{2}{*}{FM}                    & \multirow{2}{*}{98 M} & \multirow{2}{*}{100 k} & \multirow{2}{*}{B}        & \multirow{2}{*}{D, T@bistatic, L} & \multirow{2}{*}{8-element UCA}  & \multirow{2}{*}{RDA} & \multirow{2}{*}{MTT, LKF}
			{}                                                & airplane                               & 35.5 km               & 300 km                 & 5 km                                                                                                                                                                                                                                                         \\
			\cline{10-13}
			{}                                                & {}                                     & {}                    & {}                     & {}                        & {}                                & {}                              & {}                   & {}                        & small UAV                 & 45.5 km                    & 100 km                  & N/A                  \\
			\hline
			\cite{:/content/journals/10.1049/ip-rsn_20045077} & FM                                     & 96.8 M                & 1.525 k                & B                         & D, L, T@Cartesian                 & 2-element ULA                   & RDA                  & MTT, LKF                  & airplane                  & 50 km                      & 150 km                  & 10 km                \\
			\hline
			\cite{7944357}                                    & DAB                                    & 188.9 M               & 1 M                    & M (4T1R)                  & D                                 & directional antenna             & RD                   & N/A                       & fixed-wing UAV            & 11-26 km                   & 1.2 km                  & N/A                  \\
			\hline
			\cite{4721007}                                    & DAB                                    & 200 M                 & 1.5 M                  & M (3T1R)                  & D                                 & directional antenna             & RD                   & N/A                       & airplane                  & 17-20 km                   & 7-12 km                 & N/A                  \\
			\hline
			\multirow{2}{*}{\cite{4720984}}                   & DVB-T                                  & UKN                   & UKN                    & \multirow{2}{*}{M (2T1R)} & \multirow{2}{*}{D}                & \multirow{2}{*}{16-element UCA} & \multirow{2}{*}{RD}  & \multirow{2}{*}{N/A}      & \multirow{2}{*}{airplane} & \multirow{2}{*}{10-120 km} & \multirow{2}{*}{120 km} & \multirow{2}{*}{N/A} \\
			\cline{2-4}
			{}                                                & DAB                                    & 227.36 M              & 1.536 M                & {}                        & {}                                & {}                              & {}                   & {}                        & {}                        & {}                         & {}                      & {}                   \\
			\hline
			\multirow{2}{*}{\cite{8950339}}
			{}                                                & DTMB                                   & 658\&666 M            & UKN
			                                                  & \multirow{2}{*}{\tabincell{c}{M (2T1R,                                                                                                                                                                                                                                                                                                                 \\1T3R)}} & \multirow{2}{*}{D, L, T@Cartesian} & \multirow{2}{*}{UKN}     & \multirow{2}{*}{RDA}      & \multirow{2}{*}{\tabincell{c}{MTT, LKF, \\EKF}}
			                                                  & airplane                               & 20 km                 & 25 km                  & 65 m                                                                                                                                                                                                                                                         \\
			\cline{2-4}
			\cline{10-13}
			{}                                                & DTVB                                   & 658 M                 & 8 M                    &                           &                                   &                                 &                      &                           & quadcopter                & 7.5-8.3 km                 & 3 km                    & 14.51 m              \\
			\hline
			\cite{7411917}                                    & \tabincell{c}{GSM                                                                                                                                                                                                                                                                                                                                      \\(SDR-based BS)}            & 924 M                 & 200 k                         & B                               & D                                 & \tabincell{c}{8-element UCA\\w/ fixed beams}             & D                    & N/A                       & UKN                                                       & UKN         & 1 km    & N/A\\
			\hline
			\cite{9266673}                                    & GSM                                    & 945.8 M               & 81.3 k                 & M (5T1R)                  & D, T@bistatic                     & 16-element ULA                  & RDA                  & PF                        & quadcopter                & UKN                        & 320 m                   & N/A                  \\
			\hline
			\cite{7497375}                                    & GSM                                    & 1800 M                & 200 k                  & M (6T1R)                  & D                                 & 16-element ULA                  & D                    & N/A                       & quadcopter                & 0.5-16.3 km                & UKN                     & N/A                  \\
			\hline
			\cite{https://doi.org/10.1049/joe.2019.0583}      & LTE                                    & 1867.5 M              & 15 M                   & B                         & D                                 & directional antenna             & RD                   & N/A                       & quadcopter                & 50 m                       & 200 m                   & N/A                  \\
			\hline
			\cite{10.1145/3507657.3529658}                    & \tabincell{c}{LTE                                                                                                                                                                                                                                                                                                                                      \\(SDR-based BS)}            & 2495 M                & 20 M                          & M (6T1R)                        & D, L                              & directional antenna                   & RD                   & N/A                       & \tabincell{c}{quadcopter} & 1.4-8.5 m   & 20 m    & \tabincell{c}{0.515 m}\\
			\hline
			\cite{9764210}                                    & \tabincell{c}{Sensor fusion                                                                                                                                                                                                                                                                                                                            \\(LTE+GSM+camera)} & UKN                     & UKN                             & \tabincell{c}{M (1T1R LTE+\\5T1R GSM+\\4 cameras)} & D, L, T@Cartesian                 & \tabincell{c}{16-element ULA@GSM, \\8-element ULA@LTE} & RDA                  & MHT                       & quadcopter                                                & UKN           & 400 m   & 20 m\\
			\hline
			\cite{rs14236146}                                 & 5G                                     & 3440 M                & 40 M                   & B                         & D                                 & directional antenna             & RD                   & N/A                       & quadcopter                & 70 m                       & 150 m                   & N/A                  \\
			\hline
			Ours                                              & LTE                                    & 2132.5 M              & 5 M                    & B                         & D, L, T@Cartesian                 & 8-element ULA                   & RDA                  & MTT, EKF                  & quadcopter                & 255m                       & 100 m                   & 1.49 m               \\
			\hline
		\end{tabular}
	}
	\caption{An overview of the experimental works on passive UAV sensing based on broadcasting or cellular networks.
		`UKN' stands for `unknown', `N/A' stands for `not applicable', `mTnR' stands for `m transmitters and n receivers', and `R', `D' and `A' stand for `range', `Doppler' and `AoA' in the `Measurements' column, respectively.
		Note that the localization error is evaluated only for the systems with the capabilities of localization or tracking at the Cartesian domain.}
	\label{table:literature}
\end{table*}

In this work, we propose LIPASE, an LTE-based dIgital array PAssive SEnsing system, to track a target UAV by exploiting the LTE downlink signal.
Compared with the experimental results in the existing literature, LIPASE takes advantage of both larger signal bandwidth and a digital antenna array, such that the angular and range resolutions of bistatic sensing become sufficient for trajectory tracking.
Thus, the demands on out-of-band sensor or multi-static sensing can be eliminated.
The main contributions of this work are summarized as follows:
\begin{itemize}
	\item A passive sensing system with uniform linear arrays (ULAs) and digital beamforming is proposed for the detection of a target UAV, where both the bistatic range (transmitter-target-receiver range) and AoA can be measured with sufficient accuracy for trajectory tracking.
	\item With a multi-target tracking framework, it is shown that
	      the root mean squared localization error can be suppressed below 1.5 meters even with missed detections and false alarms of UAV detection.
	\item Both bistatic and Cartesian tracking methods are proposed for the multi-target tracking framework, and their performance is compared.
	      It is shown that the latter has better performance, as its state transition model could better fit the real experiment.
\end{itemize}
To the best of our knowledge, this is the first experimental work of bistatic UAV tracking with meter-level tracking accuracy.
Table \ref{table:literature} provides a comparative overview of the existing experimental results on passive UAV sensing and our proposed method.

The rest of this paper is organized as follows.
In Section \ref{sec:system}, we elaborate the system architecture and signal model.
The signal processing and trajectory tracking algorithms of the proposed LIPASE system are introduced in Sections \ref{sec:detection} and \ref{sec:tracking}, respectively.
The experimental results as well as the analysis are described in Section \ref{sec:experiment}, and the conclusion is drawn in Section \ref{sec:conclusion}.

\textit{Notation}:
Bold uppercase $\mathbf{A}$ denotes a matrix, bold lowercase $\mathbf{a}$ denotes a column vector, non-bold letters a, $A$ denote scalar values, and calligraphic letter $\mathcal{A}$ denotes sets.
$a^{*}$ denotes the conjugate of a complex value.
$[\mathbf{A}]_{i,j}$, $\mathbf{A}^{\sfT}$, and $\mathbf{A}^{\sfH}$ denote the $(i,j)$-th element, transpose, and conjugate transpose of $\mathbf{A}$, respectively.
$\mathbf{1}_{i}$ denotes the column vector whose $i$-th element is $1$ and other elements are $0$.
$\mathbb{R}^{M\!\times\!N}$ and $\mathbb{C}^{M\!\times\!N}$ denote spaces of $M\!\times\!N$ matrices with real and complex entries, respectively.
The set union, intersection, and difference operators are denoted by $\cup$, $\cap$, and $\backslash$.
The diagonal matrix constructed from a vector $\mathbf{a}$ is denoted by $\diag(\mathbf{a})$.

\section{System Overview}
\label{sec:system}
\begin{figure}[htbp]
	\centering
	\includegraphics[width=0.98\linewidth]{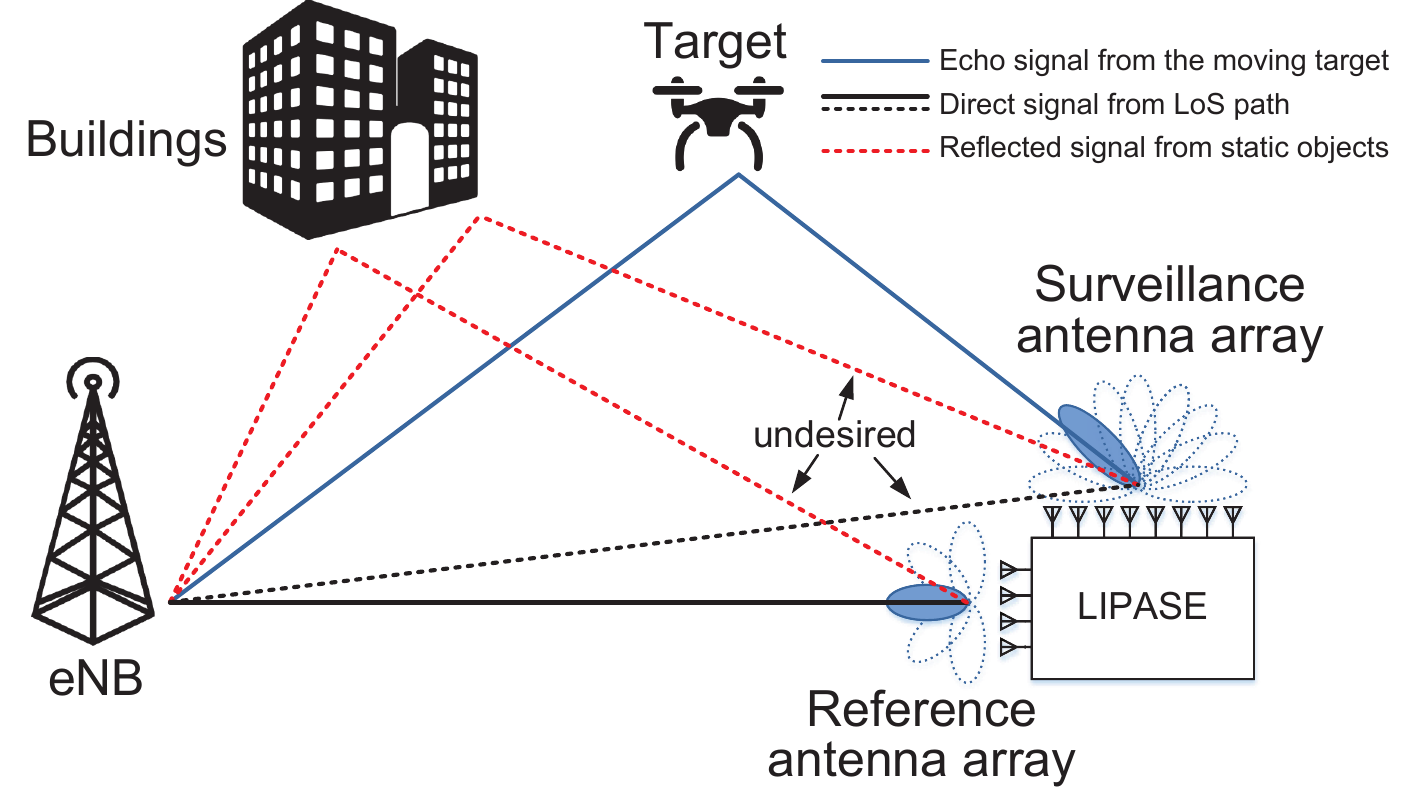}
	\caption{An example scenario of LIPASE.}
	\label{fig:system}
\end{figure}

A passive sensing system, namely LIPASE, is proposed to detect and track a target UAV by exploiting the existing cellular downlink signals in this paper.
A sensing scenario of the proposed LIPASE system is illustrated in Fig. \ref{fig:system}, which consists of an LTE eNB (Evolved Node B), a LIPASE receiver and the target UAV.
The target UAV is illuminated by the downlink signal of the eNB, such that the LIPASE receiver could detect and extract both the line-of-sight (LoS) signal from the eNB, namely the reference signal, and the scattered signal off the target UAV, namely the surveillance signal, via two antenna arrays respectively.
The bistatic Doppler shift, bistatic range and the AoA of the target UAV can then be estimated by comparing both signals, and the trajectory of the UAV can be estimated by tracking and fitting the above estimated parameters.
For elaboration convenience, the above two antenna arrays of the LIPASE receiver are referred to as the \textit{reference array} and \textit{surveillance array} respectively, and the two receive beams are referred to as the reference and surveillance beams respectively.
Moreover, both signal propagation paths are referred to as the reference and surveillance channels, respectively.

At the LIPASE receiver, both the reference and surveillance arrays follow a uniformly linear structure with $N_{\rmref}$ and $N_{\rmsur}$ patch antennas, respectively.
Due to the directional pattern of patch antennas, the eNB and the airspace to be monitored should be within the FoVs of the reference and surveillance arrays, respectively.
Note that undesired signals may also be captured by both arrays, which might interfere with the target detection.
For example, the signals scattered off the buildings in Fig. \ref{fig:system} are received by both arrays.
The receive beamforming could help to mitigate the potential interference.
On the one hand, the location of the eNB can be measured in advance, such that the reference beam can be aligned with the eNB.
On the other hand, prior knowledge on the locations of the target UAV and undesired scattering clusters is not available.
Hence, it is infeasible to direct the surveillance beam to the target UAV in advance.
Instead, the digital multiple-input multiple-output (MIMO) architecture is adopted at the LIPASE receiver, such that the directions of desired signals can be detected via multi-antenna signal processing.
As a result, the interference suppression and desired signals' extraction can be conducted in the baseband.

Particularly, let $y_{i}^{\rmref}(t)$ be the received signal at the $i$-th antenna of the reference array, $i=1,2,\ldots,N_{\rmref}$, the aggregation of the received signal at the reference array, namely the reference signal vector, can be written as $
	\mathbf{y}^{\rmref}(t)
	\triangleq
	[y_{1}^{\rmref}(t),y_{2}^{\rmref}(t),\ldots,y_{N_{\rmref}}^{\rmref}(t)]^{\sfT}
$.
It consists of the desired signal from the reference channel, $L_{\rmref}$ undesired scattered signals within the FoV, and the noise, which can be expressed by
\begin{equation}
	\label{eqn:y_ref_i}
	\mathbf{y}^{\rmref}(t)
	=
	\sum_{\ell=0}^{L_{\rmref}}a_{\ell}^{\rmref}x(t-\tau_{\ell}^{\rmref})\mathbf{a}(\phi_{\ell}^{\rmref})
	+
	\mathbf{n}^{\rmref}(t),
\end{equation}
where $x(t)$ denotes the transmit signal at the eNB,
\begin{equation}
	\label{eqn:a_phi}
	\mathbf{a}(\phi)
	\triangleq
	\left[1,e^{-j2\pi\frac{d\sin\phi}{\lambda}},\ldots,e^{-j2\pi(N_{\rmref}-1)\frac{d\sin\phi}{\lambda}}\right]^{\sfT}
\end{equation}
denotes the steering vector of the direction $\phi$, where $d$ and $\lambda$ are the antenna spacing and wavelength respectively, $\mathbf{n}^{\rmref}(t)$ denotes the noise vector, and $a_{\ell}^{\rmref}$, $\tau_{\ell}^{\rmref}$ and $\phi_{\ell}^{\rmref}$ are the attenuation coefficient, delay and AoA of the $\ell$-th signal ($\ell=0$ for the desired LoS signal) respectively.
Note that the desired reference signal comes from the LoS path, which is usually much more significant than the interference and noise.

Similarly, let $y_{i}^{\rmsur}(t)$ be the received signal at the $i$-th antenna of the surveillance array, $i=1,2,\ldots,N_{\rmsur}$.
Then the aggregation of the received signals at the surveillance array, namely the surveillance signal vector, is given by $\mathbf{y}^{\rmsur}(t)\triangleq[y_{1}^{\rmsur}(t),y_{2}^{\rmsur}(t),\ldots,y_{N_{\rmsur}}^{\rmsur}(t)]^{\sfT}$.
The surveillance signal vector consists of the desired signal from the target UAV, the undesired LoS signal, $L_{\rmsur}$ undesired scattered signals, and the noise.
It is assumed that the undesired non-line-of-sight (NLoS) signals come from static scattering clusters.
Thus, the surveillance signal vector can be written as
\begin{alignat}{2}
	 & \mathbf{y}^{\rmsur}(t)
	 &                        &
	=\sum_{\ell=0}^{L_{\rmsur}}a_{\ell}^{\rmsur}x(t-\tau_{\ell}^{\rmsur})\mathbf{a}(\phi_{\ell}^{\rmsur})                                                                                    \nonumber \\
	 &
	 &                        &
	+a^{\rmtar}x(t-\tau^{\rmtar})e^{j2\pi f^{\rmD,\rmtar}t}\mathbf{a}(\phi^{\rmtar})
	+\mathbf{n}^{\rmsur}(t),
	\label{eqn:sur}
\end{alignat}
where $a_{\ell}^{\rmsur}$, $\tau_{\ell}^{\rmsur}$ and $\phi_{\ell}^{\rmsur}$ denote the attenuation coefficient, delay and AoA of the $\ell$-th undesired signal respectively ($\ell\!=\!0$ for the undesired LoS path), $a^{\rmtar}$, $\tau^{\rmtar}$, $\phi^{\rmtar}$ and $f^{\rmD,\rmtar}$ denote the attenuation coefficient, delay, AoA and bistatic Doppler shift of the scattered signal off the target UAV respectively, and $\mathbf{n}^{\rmsur}(t)$ is the noise.
Note that the undesired LoS signal will bring significant interference to the desired surveillance signal, which should be suppressed in baseband.
Since the reference and surveillance arrays of the system are co-located, the lengths of the LoS paths from the eNB to both arrays are approximately identical, i.e., $\tau_{0}^{\rmref}\!=\!\tau_{0}^{\rmsur}\!=\!\tau_{0}$.

\section{Signal Processing for UAV Detection}
\label{sec:detection}
\begin{figure*}[htb]
	\centering
	\includegraphics[width=0.9\linewidth]{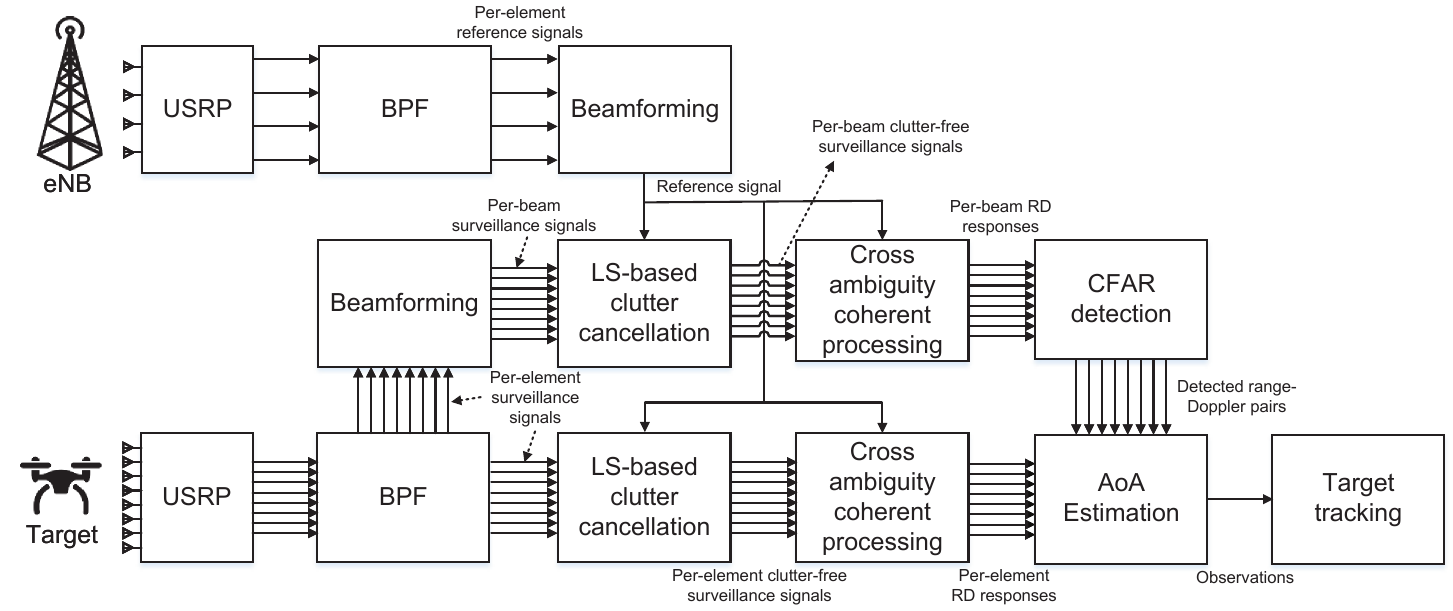}
	\caption{The proposed signal processing scheme of LIPASE.}
	\label{fig:signal_processing}
\end{figure*}

In this section, we elaborate the overall signal processing procedure of the LIPASE system for the detection of the target UAV.
As illustrated in Fig. \ref{fig:signal_processing}, the received signals at all the antenna elements of the reference and the surveillance arrays are sampled respectively at the baseband for digital processing.
In the baseband, the sampled signals of all the antennas are first filtered by band-pass filters (BPFs) to eliminate the out-of-band interference.
Then, the UAV detection consists of four steps: digital beamforming, interference cancellation, range-Doppler detection and AoA detection, which we explain one by one
in the following.

\subsection{Beamforming for Reference and Surveillance Signals}
Digital beamforming is applied to enhance the received power of both reference and surveillance signals.
Since the AoA of the reference channel is known in advance, receive beamforming along the reference channel can be applied to the reference array.
Moreover, the receive beamforming along $N_{\rmsur}$ predetermined directions is applied on the surveillance array to collect $N_{\rmsur}$ \textit{raw surveillance signals} for target detection.

Particularly, the baseband signals are sampled with a frequency $f_{\mathrm{s}}=1/T_{\mathrm{s}}$.
The sampled signals are organized by time slots each with a duration of $T$.
Thus, each time slot consists of $N=T/T_{\mathrm{s}}$ samples.
Hence, the $n$-th sampled signal vector ($\forall n=0,1,\ldots,N-1$) in the $m$-th time slot ($\forall m=1,\ldots$) at the reference and surveillance arrays can be expressed as
\begin{alignat}{3}
	 &
	\mathbf{y}^{\rmref}[m,n]
	 &   &
	=\mathbf{y}^{\rmref}\big(mT+nT_{\mathrm{s}}\big), \\
	 &
	\mathbf{y}^{\rmsur}[m,n]
	 &   &
	=\mathbf{y}^{\rmsur}\big(mT+nT_{\mathrm{s}}\big).
\end{alignat}

Let $\phi_{0}^{\rmref}$ be the AoA of the reference channel, the following beamforming is applied on the reference array:
\begin{align}
	\tilde{y}^{\rmref}[m,n]=(\mathbf{w}^{\rmref})^{\sfH}\mathbf{y}^{\rmref}[m,n],
\end{align}
where
\begin{alignat}{2}
	 &   &  &
	\mathbf{w}^{\rmref}=\mathbf{a}(\phi_{0}^{\rmref})\nonumber \\
	 & = &  &
	\left[1,e^{-j2\pi\frac{d\sin\phi_{0}^{\rmref}}{\lambda}},\ldots,e^{-j2\pi(N_{\rmref}-1)\frac{d\sin\phi_{0}^{\rmref}}{\lambda}}\right]^{\sfT}.
\end{alignat}

At the surveillance array, it might not be easy to detect the Doppler frequency or range of the target UAV directly from the received signal of one antenna element.
This is because the desired signal power is usually weak.
Since the AoA of the surveillance channel is unknown, we apply $N_{\rmsur}$ receive beams at the surveillance array along the directions
\begin{align}
	\Big\{\beta_{k}=\arcsin\Big(\frac{2(k-1)}{N_{\rmsur}}-1\Big)\Big|k=1,2,\ldots,N_{\rmsur}\Big\},
\end{align}
and the scattered signal off the target UAV is detected from the received signals after these $N_{\rmsur}$ receive beams.
Particularly, the $k$-th beamforming vector ($k\!=\!1,2,\ldots,N_{\rmsur}$) is represented by
\begin{alignat}{2}
	 &   &  &
	\mathbf{w}_{k}^{\rmsur}=\mathbf{a}(\beta_{k})\nonumber \\
	 & = &  &
	\left[1,e^{-j2\pi\frac{d\sin\beta_{k}}{\lambda}},\ldots,e^{-j2\pi(N_{\rmref}-1)\frac{d\sin\beta_{k}}{\lambda}}\right]^{\sfT}.
\end{alignat}
The signal after beamforming vector $\mathbf{w}_{k}$, namely the $k$-th \textit{raw surveillance signal}, is given by
\begin{align}
	\tilde{y}_{k}^{\rmsur}[m,n]=(\mathbf{w}_{k}^{\rmsur})^{\sfH}\mathbf{y}^{\rmsur}[m,n].
\end{align}

\subsection{Interference Cancellation of Raw Surveillance Signals}
\label{subsec:cluterCancellation}
The received signal of the surveillance array in \eqref{eqn:sur} may suffer from strong interference.
For example, the LoS signal power from the serving eNB is usually much larger than the desired signal power at the surveillance array.
The scattered signals off non-target clusters may also be strong at the surveillance array.
With the assistance of the reference signal, the technique of clutter cancellation (CC) can be applied to suppress the above interference, which is usually with a zero or low Doppler frequency.

Particularly, the least-squares-based (LS-based) CC algorithm \cite{1631830} is applied on each antenna of the surveillance array, as well as the $N_{\rmsur}$ raw surveillance signals, respectively.
The latter will be used to detect the Doppler frequency and range of the target UAV, and the former will be used to detect the AoA of the surveillance channel.
Without loss of generality, the interference suppression on the $k$-th raw surveillance signal ($k=1,2,\ldots,N_{\rmsur}$) is as given below.
The interference suppression on each antenna element can be applied similarly.
We first define sampled signal vectors in the $m$-th time slot as
\begin{align}
	\tilde{\bm{\omega}}_{m}^{\rmsur}
	=
	\big[
		\tilde{y}_{k}^{\rmsur}[m,0],
		\tilde{y}_{k}^{\rmsur}[m,1],
		\ldots,
		\tilde{y}_{k}^{\rmsur}[m,N-1]
		\big]^{\sfT},
\end{align}
and
\begin{align}
	\bm{\omega}_{m}^{\rmref\!,l}
	\!=\!
	\big[
		\tilde{y}^{\rmref}[m,\!-\!l],
		\tilde{y}^{\rmref}[m,\!-\!l\!+\!1],
		\ldots,
		\tilde{y}^{\rmref}[m,\!-\!l\!+\!N\!-\!1]
		\big]^{\sfT},\nonumber
\end{align}
where $\tilde{y}^{\rmref}[m,-l]=\tilde{y}^{\rmref}[m-1,-l+N-1]$ represents the $l$-th sample before the $m$-th time slot.
In other words, it is the $(-l+N-1)$-th sample in the $(m-1)$-th time slot.

Let $T=N/f_{\mathrm{s}}$ be the duration of coherent integration time (CIT), which is the time span for the Doppler frequency detection.
The detection resolution of Doppler frequency is $f_{\mathrm{s}}/N$.
The CC algorithm is adopted to suppress the interference with the Doppler frequencies in $[-Pf_{\mathrm{s}}/N,Pf_{\mathrm{s}}/N]$ and the delay in $[-L/f_{\mathrm{s}},L/f_{\mathrm{s}}]$, where $P$ and $L$ are two constants.
As a remark, note that $Pf_{\mathrm{s}}/N$ should be smaller than the possible Doppler frequency of the target UAV.

Let $
	\bm{\Omega}_{m}^{\rmref}
	\in
	\mathbb{C}^{N\times(2P+1)(L+1)}
$ be a matrix aggregating the reference signals with different delays and Doppler frequencies.
Particularly, its $\big((p\!+\!P)(L\!+\!1)\!+\!l\!+\!1\big)$-th column is the $l$-sample-delayed and $p$-Doppler-shifted version ($0\!\leq\! l\!\leq\! L,-\!P\!\leq\! p\!\leq\! P$) of the reference signal as
\begin{alignat}{2}
	 &   &  &
	[\bm{\Omega}_{m}^{\rmref}]_{(p+P)(L+1)+l+1}\nonumber \\
	 & = &  &
	\mathbf{Diag}(1,e^{j2\pi p/N},\ldots,e^{j2\pi p(N-1)/N})\bm{\omega}_{m}^{\rmref,l}.
\end{alignat}
The LS-based CC algorithm is to find a cancellation filter $\mathbf{h}_{m}^{\star}\!\in\!\mathbb{C}^{(2P+1)(L+1)\times 1}$ to minimize the residual signal power after cancellation.
Thus,
\begin{align}
	\label{eqn:LS}
	\mathbf{h}_{m}^{\star}
	=
	\argmin_{\mathbf{h}_{m}}
	\big\|\tilde{\bm{\omega}}_{m}^{\rmsur}-\bm{\Omega}_{m}^{\rmref}\mathbf{h}_{m}\big\|^{2},
\end{align}
whose optimal solution can be represented by
\begin{align}
	\mathbf{h}_{m}^{\star}
	=
	([\bm{\Omega}_{m}^{\rmref}]^{\sfH}\bm{\Omega}_{m}^{\rmref})^{-1}[\bm{\Omega}_{m}^{\rmref}]^{\sfH}\tilde{\bm{\omega}}_{m}^{\rmsur}.
\end{align}

Therefore, the $k$-th raw surveillance signal after interference suppression ($\forall k\!=\!1,2,\ldots,N_{\rmsur}$) is given by the projection of the raw surveillance signal onto the subspace orthogonal to the clutter subspace, i.e.,
\begin{align}
	\label{eqn:CC}
	\tilde{\bm{\omega}}_{m}^{\rmsur,\rmcf}
	=\tilde{\bm{\omega}}_{m}^{\rmsur}-\bm{\Omega}_{m}^{\rmref}\mathbf{h}_{m}^{\star}
	={\bf\Phi}_{m}\tilde{\bm{\omega}}_{m}^{\rmsur},
\end{align}
where
\begin{align}
	{\bf\Phi}_{m}
	=\mathbf{I}-\bm{\Omega}_{m}^{\rmref}([\bm{\Omega}_{m}^{\rmref}]^{\sfH}\bm{\Omega}_{m}^{\rmref})^{-1}[\bm{\Omega}_{m}^{\rmref}]^{\sfH}.
\end{align}

In the remainder of this paper, the $N_{\rmsur}$ raw surveillance signals and the $N_{\rmsur}$ signals on each antenna element after interference suppression are denoted as
\begin{align}
	\tilde{y}_{k}^{\rmsur,\rmcf}[m,n]\triangleq[\tilde{\bm{\omega}}_{m}^{\rmsur,\rmcf}]_{n}
\end{align}
and
\begin{align}
	y_{i}^{\rmsur,\rmcf}[m,n]\triangleq[\bm{\omega}_{m,i}^{\rmsur,\rmcf}]_{n},
\end{align}
$\forall m,n$, respectively, where $[\cdot]_n$ denotes the $n$-th entry of the vector.

\subsection{Range-Doppler Detection}
The \textit{range-Doppler (RD) response} of the target UAV can be detected from the cross ambiguity function (CAF) of the $N_{\rmsur}$ raw surveillance signals.
Without loss of generality, the range-Doppler detection in the $m$-th time slot is elaborated in this part.
First, the CAF of the $k$-th raw surveillance signal ($\forall k$) in the $m$-th time slot, denoted by $\tilde{A}_{m,k}\in\mathbb{C}^{(L+1)\times N}$, is defined as
\begin{align}
	\label{eqn:B}
	\tilde{A}_{m,k}[l,p]
	 &
	=
	\sum_{n=0}^{N-1}\tilde{y}_{k}^{\rmsur,\rmcf}[m,n]\big[\tilde{y}^{\rmref}[m,n-l]\big]^{*}e^{-j2\pi pn/N},\nonumber \\
	 &
	\ \forall \ 0 \leq l \leq L,\ -\frac{N}{2}\leq p<\frac{N}{2}.
\end{align}
A target candidate with bistatic Doppler frequency $pf_{\mathrm{s}}/N$ at a bistatic range difference (the difference between the transmitter-target-receiver range and transmitter-receiver range) $l c/f_{\mathrm{s}}$ can be detected if $\tilde{A}_{m,k}$ has a clear peak value at the $(l,p)$-th entry.

A two-dimensional cell-averaging constant false alarm rate (CA-CFAR) detector with guard cells \cite{:/content/journals/10.1049/ip-rsn_20045077} and detection clustering \cite{10005216} is applied for target detection in the $N_{\rmsur}$ raw surveillance signals respectively.
First, for each cell in an RD response, the average noise power is estimated by its surrounding cells.
Denote $L_{\rmguard}$ and $P_{\rmguard}$ as the number of guard range and Doppler bins respectively, $L_{\rmtrain}$ and $P_{\rmtrain}$ as the number of training range and Doppler bins respectively, the noise power at the $(l,p)$-th cell is estimated by
\begin{equation}
	P_{m,k}^{l,p}=\frac{1}{N_{\rmtrain}}\sum_{(l',p')\in\mathcal{W}_{l,p}^{\rmtotal}\backslash\mathcal{W}_{l,p}^{\rmguard}}\big|\tilde{A}_{m,k}[l',p']\big|,
\end{equation}
where
\begin{align*}
	\mathcal{W}_{l,p}^{\rmtotal}
	=
	\big\{(l',p')\big|
	 & |l'-l|\leq L_{\rmguard}+L_{\rmtrain},       \\
	 & |p'-p|\leq P_{\rmguard}+P_{\rmtrain}\big\},
\end{align*}
\begin{align*}
	\mathcal{W}_{l,p}^{\rmguard}
	=
	\big\{(l',p')\big||l'-l|\leq L_{\rmguard},
	|p'-p|\leq P_{\rmguard}\big\},
\end{align*}
and
$N_{\rmtrain}= |\mathcal{W}_{l,p}^{\rmtotal}| -|\mathcal{W}_{l,p}^{\rmguard}|$.
Then, a target candidate is detected at the $(l,p)$-th cell of the RD response, if
\begin{align*}
	\big|\tilde{A}_{m,k}[l,p]\big|\geq \alpha P_{m,k}^{l,p},
\end{align*}
where $\alpha$ denotes the threshold factor.

As a result, the set of detected target candidates in all the $N_{\rmsur}$ raw surveillance signals and the $m$-th time slot ($\forall m$) is denoted as
\begin{eqnarray}
	\mathcal{P}_{m}
	&\triangleq&
	\Big\{(l,p)\Big|\big|\tilde{A}_{m,k}[l,p]\big|\geq \alpha P_{m,k}^{l,p},\forall k\Big\}.
\end{eqnarray}

\subsection{AoA Estimation}
\label{subsec:target_detection}
Given the detected target candidates $\mathcal{P}_{m}$ in the $m$-th time slot ($\forall m$), the detection of the corresponding AoAs is elaborated in this part.
For notation convenience, denote the $u$-th pair of the set $\mathcal{P}_{m}$ as $(l_{m,u},p_{m,u})$.

In order to detect the AoAs, we first calculate the CAF for all the detected target candidates in $\mathcal{P}_{m}$ at all the antenna elements.
Particularly, the RD responses of the $i$-th surveillance antenna element in the $m$-th time slot can be calculated by
\begin{align}
	\label{eqn:A}
	 & A_{m,i}[l_{m,u},p_{m,u}]=\sum_{n=0}^{N-1}\Big\{y_{i}^{\rmsur,\rmcf}[m,n]\big[\tilde{y}^{\rmref}[m,n-l_{m,u}]\big]^{*}\nonumber \\
	 & \times e^{-j2\pi p_{m,u}n/N}\Big\},\forall m,i,u,\ (l_{m,u},p_{m,u})\in \mathcal{P}_{m}.
\end{align}
Then, the vector of RD response for all pairs of bistatic range and Doppler frequency in $\mathcal{P}_{m}$ is defined as
\begin{align}
	\mathbf{a}_{m,u}=\big[A_{m,1}[l_{m,u},p_{m,u}],\ldots,\allowbreak A_{m,N_{\rmsur}}[l_{m,u},p_{m,u}]\big]^{\sfT}.
\end{align}

Using the phase interferometry method \cite{:/content/journals/10.1049/ip-rsn_20045077}, the AoA of the surveillance channel with respect to the broadside direction of the antenna array can be estimated by
\begin{align}
	\hat{\theta}_{m,u}=\argmax_{\theta}\left|\mathbf{a}_{m,u}^{\sfH}\mathbf{a}(\theta)\right|,\forall u=1,2,\ldots,|\mathcal{P}_{m}|.
\end{align}
The solution can be derived by linear search on fine-grained directions.
Let $\Delta\theta_{\rmsur}$ be the broadside direction of the surveillance array versus the x-axis as illustrated in Fig. \ref{fig:measurementFunction}, the directions of the target candidates in $\mathcal{P}_{m}$ versus the x-axis are represented by
\begin{align}
	\theta_{m}^{u}\triangleq\Delta\theta_{\rmsur}+\hat{\theta}_{m,u}.
\end{align}

\subsection{Set of Observations}
Denote the locations of the eNB and the LIPASE system as $\mathbf{l}_{\rmtx}\triangleq[x_{\rmtx},y_{\rmtx}]^{\sfT}$ and $\mathbf{l}_{\rmrx}\triangleq[x_{\rmrx},y_{\rmrx}]^{\sfT}$, respectively.
The baseline range, i.e., the distance between the eNB and LIPASE receiver, is given by
\begin{align}
	L_{\rmtxrx}=\|\mathbf{l}_{\rmtx}-\mathbf{l}_{\rmrx}\|.
\end{align}
Then the bistatic range of the $u$-th detected target candidate in $\mathcal{P}_{m}$, i.e., the transmitter-target-receiver range, is given by
\begin{align}
	R_{m}^{u}=l_{m,u}c/f_{\mathrm{s}}+L_{\rmtxrx}.
\end{align}
Moreover, the corresponding bistatic range rate is given by
\begin{align}
	\dot{R}_{m}^{u}=-\lambda p_{m,u}f_{\mathrm{s}}/N.
\end{align}

As a result, aggregating the measured bistatic range, range rate and AoA of target candidates in $\mathcal{P}_{m}$, the set of detected target candidates in the $m$-th time slot can be rewritten as
\begin{align}
	\mathcal{D}_{m}\triangleq\{\mathbf{z}_{m}^{u}|u=1,2,\ldots,|\mathcal{P}_{m}|\},
\end{align}
where $\mathbf{z}_{m}^{u}\triangleq[R_{m}^{u},\dot{R}_{m}^{u},\theta_{m}^{u}]^{\sfT}$ is referred to as the $u$-th observation in $\mathcal{D}_{m}$.

\section{Target Tracking Scheme}
\label{sec:tracking}
Due to the presence of false alarms and missed detections, there might be zero or multiple observations in the passive sensing of one time slot.
The set of observations, $\mathcal{D}_{m}$ ($\forall m=1,2,\ldots$), cannot be directly used as the trajectory of the target UAV.
Even with a single observation in each slot, the detection error might be large due to the limited signal bandwidth and antenna number (limited range and angular resolutions).
Note that the velocity of the UAV can be partially observed from the bistatic geometry and Doppler frequency, the past observations of position and velocity can be exploited in the localization of the target UAV in each time slot.
Moreover, because of potential false alarms in the detections, the multi-target tracking (MTT) algorithm \cite{blackman1999design} is utilized for the trajectory tracking.
Hence, multiple tracks of the trajectory can be dynamically established, updated, or eliminated according to the observations in each time slot, and the survival track at the end will be considered the best estimation of the trajectory.

The MTT algorithm tracks the variation of the target state, and two different definitions of target state are considered: (1) the target state is represented by the bistatic range, Doppler frequency and AoA; and (2) the target state is represented by its position and velocity.
The former and the latter are referred to as the \textit{bistatic tracking} and \textit{Cartesian tracking}, respectively, for elaboration convenience.
Note that either tracking method is sufficient to deduce the trajectory of the target UAV; their performance will be compared and discussed in Section \ref{sec:experiment}.
Let $\mathbf{s}_{m}$, $m=1,2,\ldots$, be the state of the target UAV in the $m$-th time slot.
When the bistatic tracking is considered, the state can be expressed as
\begin{equation}
	\mathbf{s}_{m} =[R_{m},\dot{R}_{m},\ddot{R}_{m},\theta_{m},\dot{\theta}_{m}]^{\sfT},
\end{equation}
which consists of the bistatic range $R_{m}$, AoA $\theta_{m}$, and their derivatives $\dot{R}_{m},\ddot{R}_{m}, \dot{\theta}_{m}$ with respect to time.
On the other hand, when the Cartesian tracking is considered, the state can be expressed as
\begin{equation}
	\mathbf{s}_{m} = [x_{m},\dot{x}_{m},y_{m},\dot{y}_{m}]^{\sfT},
\end{equation}
which consists of the coordinates $\mathbf{l}_{m}=[x_{m},y_{m}]^T$ and velocity $\mathbf{v}_{m}=[\dot{x}_{m},\dot{y}_{m}]^T$ of the target UAV.

\begin{figure}[htb]
	\centering
	\includegraphics[width=0.6\linewidth]{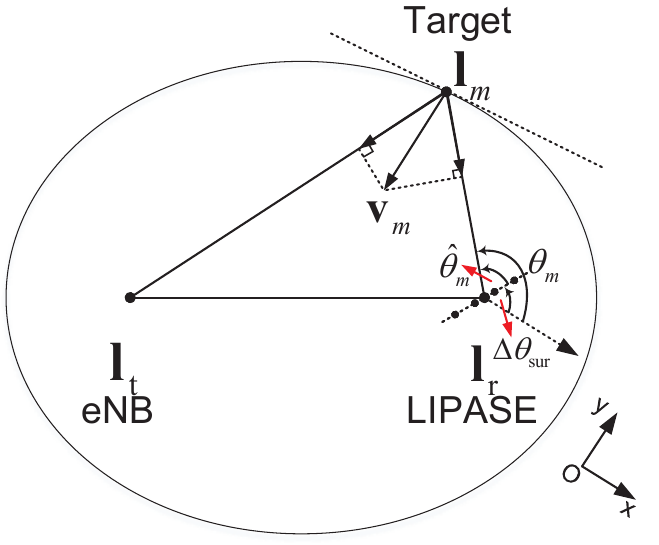}
	\caption{Geometry of passive sensing.}
	\label{fig:measurementFunction}
\end{figure}

Moreover, the measurement model, which maps the target state to the passive sensing observation, is given by
\begin{equation}
	\label{eqn:measurementModel}
	\mathbf{z}_{m}=\mathbf{h}(\mathbf{s}_{m})+\mathbf{u}_{m},
\end{equation}
where $\mathbf{h}(\cdot)$ is the measurement function, $\mathbf{u}_{m}\sim\mathcal{N}(\mathbf{0},\mathbf{R})$ denotes the measurement noise with covariance matrix $\mathbf{R}$.
When the bistatic tracking is considered, it is clear that the measurement function can be rewritten as
\begin{equation}
	\label{eqn:measurementFunctionBistatic}
	\mathbf{h}(\mathbf{s}_{m})=\mathbf{H}\mathbf{s}_{m},
\end{equation}
where the measurement matrix $\mathbf{H}$ is defined as
\begin{equation}
	\label{eqn:H_bist}
	\mathbf{H}
	\triangleq
	\left[\begin{array}{ccccc}
			1 & 0 & 0 & 0 & 0 \\
			0 & 1 & 0 & 0 & 0 \\
			0 & 0 & 0 & 1 & 0 \\
		\end{array}\right].
\end{equation}
On the other hand, when the Cartesian tracking is considered, the nonlinear measurement function can be derived according to the geometry in Fig. \ref{fig:measurementFunction} as
\begin{equation}
	\label{eqn:measurementFunctionCartesian}
	\mathbf{h}(\mathbf{s}_{m})
	=
	\left[
		\begin{array}{c}
			\|\mathbf{l}_{m}-\mathbf{l}_{\rmrx}\|
			+
			\|\mathbf{l}_{m}-\mathbf{l}_{\rmtx}\|
			\\
			\left[
				\frac{
					\mathbf{l}_{m}-\mathbf{l}_{\rmrx}}{
					\|\mathbf{l}_{m}-\mathbf{l}_{\rmrx}\|}
				+
				\frac{
					\mathbf{l}_{m}-\mathbf{l}_{\rmtx}}{
					\|\mathbf{l}_{m}-\mathbf{l}_{\rmtx}\|}
				\right]^{\sfT}
			\cdot
			\mathbf{v}_{m}
			\\
			\atantwo\left(y_{m}-y_{\rmrx},x_{m}-x_{\rmrx}\right)
			\\
		\end{array}
		\right],
\end{equation}
where $\atantwo(\cdot,\cdot)$ denotes a four-quadrant inverse tangent.

In the following, we first introduce the procedure of a multi-target tracking algorithm with general operators of track initialization, prediction, and update.
Then, these operators are specified for both definitions of the UAV state.

\subsection{The Multi-Target Tracking Framework}
\label{subsec:MTT}
The MTT algorithm initializes and maintains a set of tracks, where each track consists of a sequence of estimated UAV states and the corresponding covariance matrices of estimation errors.
In each time slot, the tracks are updated according to the latest observations from passive sensing.

Particularly, denote the set of tracks obtained in the $m$-th time slot as $\mathcal{T}_{m}$, $\forall m$.
They are further divided into two subsets: \textit{tentative track} (TT) subset $\mathcal{T}_{m}^{\rmtt}$ and \textit{confirmed track} (CT) subset $\mathcal{T}_{m}^{\rmct}$, and thus $\mathcal{T}_{m}=\mathcal{T}_{m}^{\rmtt}\cup\mathcal{T}_{m}^{\rmct}$.
Initially, the confirmed track subset is empty, i.e.,
\begin{equation}
	\mathcal{T}_{1}^{\rmct}=\emptyset,
\end{equation}
and the tentative track subset is given by
\begin{equation}
	\mathcal{T}_{1}^{\rmtt} = \{\tau_{1}^{u}=(\mathbf{s}_{1|1}^{u},\mathbf{P}_{1|1}^{u}) |\forall  \mathbf{z}_{1}^{u}\in\mathcal{D}_{1}\},
\end{equation}
where $\tau_{1}^{u}$ is the $u$-th track in the first time slot, $\mathbf{s}_{1|1}^{u}$ and $\mathbf{P}_{1|1}^{u}$ denote the $u$-th estimation of $\mathbf{s}_{1}$ and the covariance matrix of estimation error according to the $u$-th observation $\mathbf{z}_{1}^{u}$ in $\mathcal{D}_{1}$.
The track initialization of $(\mathbf{s}_{1|1}^{u},\mathbf{P}_{1|1}^{u})$ can be represented by the operator
\begin{equation}
	\label{eqn:initTrack1}
	(\mathbf{s}_{1|1}^{u},\mathbf{P}_{1|1}^{u})=\textrm{InitTrack}(\mathbf{z}_{1}^{u}),
\end{equation}
which will be further explained in Section \ref{sec:tracking}-\ref{subsec:init}.
Then, the update of the tracks in the $m$-th time slot, $m=2,3,\ldots$, is elaborated below.
The whole algorithm is also summarized in Algorithm \ref{alg:MTT}.

{\bf Step 1: State Prediction.} The UAV state of the $m$-th time slot is predicted according to the tracks estimated in the $(m-1)$-th time slot.
Particularly, all the tracks in $\mathcal{T}_{m-1}$ are assigned with unique indexes, and each track (say the $v$-th track $\tau_{m-1}^{v}$) is used to predict a possible UAV state, denoted as $\mathbf{s}_{m|m-1}^{v}$.
The covariance matrix of prediction error $\mathbf{P}_{m|m-1}^{v}$ of the $m$-th time slot is also calculated.
The prediction can be expressed by the operator
\begin{equation}
	\label{eqn:predictTrack}
	(\mathbf{s}_{m|m-1}^{v},\mathbf{P}_{m|m-1}^{v})=\textrm{PredictTrack}(\tau_{m-1}^{v}), \forall \tau_{m-1}^{v}\in \mathcal{T}_{m-1},
\end{equation}
which will be further elaborated in Section \ref{sec:tracking}-\ref{subsec:trackPrediction}.

{\bf Step 2: Observation-to-Track Association.} Each observation in $\mathcal{D}_{m}$ is associated with the closest track in $\mathcal{T}_{m-1}$ with the gating constraint.
We first define the Mahalanobis distance between an arbitrary observation $\mathbf{z}_{m}^{u} \in \mathcal{D}_{m}$ and a state prediction tuple $(\mathbf{s}_{m|m-1}^{v},\mathbf{P}_{m|m-1}^{v})$ as
\begin{equation}
	d_{\rmMah}(\mathbf{z}_{m}^{u};\mathbf{s}_{m|m-1}^{v},\mathbf{P}_{m|m-1}^{v})\triangleq\sqrt{[\Delta\mathbf{z}_{m}^{u,v}]^{\sfT}[\mathbf{W}_{m}^{v}]^{-1}\Delta\mathbf{z}_{m}^{u,v}},\nonumber
\end{equation}
where
\begin{equation}
	\Delta\mathbf{z}_{m}^{u,v}\triangleq\mathbf{z}_{m}^{u}-\mathbf{h}(\mathbf{s}_{m|m-1}^{v}) \nonumber
\end{equation}
denotes the measurement residual,
\begin{equation}
	\mathbf{W}_{m}^{v}\triangleq\mathbf{R}+\mathbf{H}_{m}^{v}\mathbf{P}_{m|m-1}^{v}[\mathbf{H}_{m}^{v}]^{\sfT} \nonumber
\end{equation}
denotes the covariance matrix of measurement residual, and
\begin{equation}
	\label{eqn:JacobianMeasurementMatrix}
	\mathbf{H}_{m}^{v}
	\triangleq
	\mathbf{J}_{\mathbf{h}}|_{\mathbf{s}=\mathbf{s}_{m|m-1}^{v}}
	=
	\left.\frac{\partial \mathbf{h}(\mathbf{s})}{\partial \mathbf{s}}\right|_{\mathbf{s}=\mathbf{s}_{m|m-1}^{v}}
\end{equation}
denotes the Jacobian of the measurement function given the predicted state of track $\tau_{m-1}^{v}$.
The following gating constraint should be satisfied in each association:
\begin{align}
	\label{eqn:gating}
	d_{\rmMah}^{2}(\mathbf{z}_{m}^{u};\mathbf{s}_{m|m-1}^{v},\mathbf{P}_{m|m-1}^{v})\leq \gamma,\ \forall u,v,
\end{align}
where $\gamma$ is a threshold.

Then, we define the cost associating the observation $\mathbf{z}_{m}^{u}$ with the state prediction (or track equivalently) $\mathbf{s}_{m|m-1}^{v}$ as the sum of the squared Mahalanobis distance and a penalty for large prediction covariance.
Thus,
\begin{align}
	\label{eqn:dist-const}
	C_{m}^{u,v}\triangleq d_{\rmMah}^{2}(\mathbf{z}_{m}^{u};\mathbf{s}_{m|m-1}^{v},\mathbf{P}_{m|m-1}^{v})+\ln(|\mathbf{W}_{m}^{v}|),
\end{align}
where $|\mathbf{W}_{m}^{v}|$ denotes the determinant of matrix $\mathbf{W}_{m}^{v}$.

Let $a_{u,v}\in\{0,1\}$ be the indicator of association, whose value is $1$ when the observation $\mathbf{z}_{m}^{u}$ is associated with the state prediction $\mathbf{s}_{m|m-1}^{v}$ (equivalently the track $\tau_{m-1}^{v}$) and $0$ otherwise.
The observation-to-track association consists of two steps.
In the first step, the observations in $\mathcal{D}_{m}$ are associated with the tracks in $\mathcal{T}_{m-1}^{\rmct}$ as
\begin{subequations}
	\begin{alignat}{3}
		\textsf{P1:}\
		 & \{a_{m}^{u,v,\star}\}=
		 &                        & \argmin_{\{a_{m}^{u,v}\}}
		\sum_{u:\mathbf{z}_{m}^{u}\in\mathcal{D}_{m}}\sum_{v:\tau^{v}_{m-1}\in \mathcal{T}_{m-1}^{\rmct}}a_{m}^{u,v}C_{m}^{u,v},\nonumber
		\\
		\label{eqn:GNNCT2}
		 &                        &                           & \mathrm{s.t.}
		\sum_{u:\mathbf{z}_{m}^{u}\in\mathcal{D}_{m}}a_{m}^{u,v}\leq 1,
		\forall v,
		\\
		\label{eqn:GNNCT3}
		 &                        &                           & \hspace{2ex}                       \sum_{v:\tau_{m-1}^{v}\in \mathcal{T}_{m-1}^{\rmct}} a_{m}^{u,v}\leq 1,
		\forall u,
		\\
		 &                        &                           & \hspace{4ex} \mbox{Gating constraint in \eqref{eqn:gating}},\nonumber
	\end{alignat}
\end{subequations}
where \eqref{eqn:GNNCT2} ensures that each track can be associated with at most one observation, and vice versa in \eqref{eqn:GNNCT3}.
The optimal solution of the association problem can be solved by the Hungarian algorithm \cite{blackman1999design}, and the set of the associated observations in the above optimization is denoted as
\begin{equation}
	\mathcal{A}_{m}^{\rmct} = \{\mathbf{z}_{m}^{u}|\forall u,a_{m}^{u,v,\star}=1,\tau_{m-1}^{v}\in\mathcal{T}_{m-1}^{\rmct}\}.
\end{equation}

Let $\widetilde{\mathcal{D}}_{m}^{\rmrem}=\mathcal{D}_{m}\backslash\mathcal{A}_{m}^{\rmct}$ be the set of remaining observations without association; it is further considered in the association with $\mathcal{T}_{m-1}^{\rmtt}$ as
\begin{subequations}
	\begin{alignat}{3}
		\textsf{P2:}\
		 & \{a_{m}^{u,v,\star}\}=
		 &                        & \argmin_{\{a_{m}^{u,v}\}}
		\sum_{u:\mathbf{z}_{m}^{u}\in\widetilde{\mathcal{D}}_{m}^{\rmrem}}\sum_{v:\tau^{v}_{m-1}\in\mathcal{T}_{m-1}^{\rmtt}} a_{m}^{u,v}C_{m}^{u,v},\nonumber
		\\
		\label{eqn:GNNTT2}
		 &                        &                           & \mathrm{s.t.}
		\sum_{u:\mathbf{z}_{m}^{u}\in\widetilde{\mathcal{D}}_{m}^{\rmrem}}a_{m}^{u,v}\leq 1,
		\forall v,
		\\
		\label{eqn:GNNTT3}
		 &                        &                           & \hspace{2ex}                                                                                                                                                                                  \sum_{v:\tau^{v}_{m-1}\in \mathcal{T}_{m-1}^{\rmtt}} a_{m}^{u,v}\leq 1,
		\forall u,                                                                                                                                                                                                                                                                                                                    \\
		 &                        &                           & \hspace{4ex}                \mbox{Gating constraint in \eqref{eqn:gating}}.\nonumber
	\end{alignat}
\end{subequations}
Denote the set of associated observations in the above optimization $\textsf{P2}$ as
\begin{equation}
	\mathcal{A}_{m}^{\rmtt}=\{\mathbf{z}_{m}^{u}| \forall u,a_{m}^{u,v,\star}=1,\tau^{v}_{m-1}\in\mathcal{T}_{m-1}^{\rmtt}\}.
\end{equation}
Let $\mathcal{A}_{m}=\mathcal{A}_{m}^{\rmct} \cup \mathcal{A}_{m}^{\rmtt}$, then the set of remaining observations without association can be represented as
\begin{equation}
	\mathcal{D}_{m}^{\rmrem}=\widetilde{\mathcal{D}}_{m}^{\rmrem}\backslash\mathcal{A}_{m}^{\rmtt}=\mathcal{D}_{m}\backslash\mathcal{A}_{m}.
\end{equation}

{\bf Step 3: Track Maintenance.} This step includes the track initialization, confirmation, and deletion.
First, a tentative track is confirmed as a confirmed track if it has been associated with an observation in each of the latest $N^{\rmcnf}$ time slots.
Then, a track will be removed from the tentative track set or the confirmed track set, if it has not been associated with any observation in any of the latest $N^{\rmdel}$ time slots.
Denote the confirmed track set and the tentative track set after the above confirmation and deletion as $\mathcal{T}_{m-1}^{\rmtt,\prime}$ and $\mathcal{T}_{m-1}^{\rmct,\prime}$, respectively.
Finally, the observations in $\mathcal{D}_{m}^{\rmrem}$ are initialized as new tracks, i.e.,
\begin{align}
	\label{eqn:initTrackM}
	\mathcal{T}_{m}^{\rminit}=\{ & (\mathbf{s}_{m|m}^{u},\mathbf{P}_{m|m}^{u})\nonumber                                                \\
	                             & =\textrm{InitTrack}(\mathbf{z}_{m}^{u})|\forall u,\mathbf{z}_{m}^{u}\in\mathcal{D}_{m}^{\rmrem} \}.
\end{align}

{\bf Step 4: Track Update.}
The tracks in $\mathcal{T}_{m-1}^{\rmtt,\prime}$ and $\mathcal{T}_{m-1}^{\rmct,\prime}$ are updated according to the associated observations or prediction (if no observation is associated with) in this step.
Suppose the $v$-th track in $\mathcal{T}_{m-1}$, denoted as $\tau_{m-1}^{v}$, is also in $\mathcal{T}_{m-1}^{\rmtt,\prime} \cup \mathcal{T}_{m-1}^{\rmct,\prime}$.
The state of the target UAV in the $m$-th time slot along this track, as well as the covariance matrix of estimation error, can be obtained as
\begin{fleqn}
	\begin{align*}
		(\mathbf{s}_{m|m}^{v},\mathbf{P}_{m|m}^{v})=
	\end{align*}
\end{fleqn}
\begin{subnumcases}{}
	\label{eqn:updateTrackA}
	\textrm{CorrectTrack}(\tau_{m-1}^{v},\mathbf{z}_{m}^{u}), & $\exists u, a_{m}^{u,v,\star}=1$\\
	\label{eqn:updateTrackB}
	(\mathbf{s}_{m|m-1}^{v},\mathbf{P}_{m|m-1}^{v}), & $\textrm{otherwise}$
\end{subnumcases}
where \eqref{eqn:updateTrackA} refers to the situation that there is one observation in $\mathcal{D}_{m}$ associated with this track, and \eqref{eqn:updateTrackB} refers to otherwise.
The track correction operator $\textrm{CorrectTrack}(\cdot)$ for the two tracking methods will be further elaborated in Section \ref{sec:tracking}-\ref{subsec:trackCorrection}.
Finally, the tentative and confirmed track subsets are updated as
\begin{alignat}{2}
	\label{eqn:updateCT}
	 & \mathcal{T}_{m}^{\rmct}
	 &                         & =\{\tau_{m}^{v}\triangleq(\mathbf{s}_{m|m}^{v},\mathbf{P}_{m|m}^{v},\tau_{m-1}^{v})|\forall \tau_{m-1}^{v}\in \mathcal{T}_{m-1}^{\rmct,\prime}\}                                       \\
	\label{eqn:updateTT}
	 & \mathcal{T}_{m}^{\rmtt}
	 &                         & =\{\tau_{m}^{v}\triangleq(\mathbf{s}_{m|m}^{v},\mathbf{P}_{m|m}^{v},\tau_{m-1}^{v})|\forall \tau_{m-1}^{v}\in \mathcal{T}_{m-1}^{\rmtt,\prime}\} \cup \mathcal{T}_{m}^{\textrm{init}}.
\end{alignat}

\begin{algorithm}[htb]
	\DontPrintSemicolon
	\SetInd{0.2em}{0.7em}
	\SetKwProg{ForEachTimeSlot}{At the end of the $m$-th time slot ($m=2,3,\ldots$), obtain $\mathcal{D}_{m}$:}{}{end}
	\SetKwProg{ForEachTrackA}{For track $\tau_{m-1}^{v}$ ($\forall v\in\mathcal{T}_{m-1}$):}{}{end}
	\SetKwProg{ForEachTrackB}{For track $\tau_{m-1}^{v}$ ($\forall v\in\mathcal{T}_{m-1}^{\rmct,\prime}\cup\mathcal{T}_{m-1}^{\rmtt,\prime}$):}{}{end}
	\SetKwProg{ForEachDetectionB}{For the $u$-th detection ($\forall\mathbf{z}_{m}^{u}\in\mathcal{D}_{m}^{\rmrem}$):}{}{end}
	\BlankLine

	\textbf{At the end of the $1$-st time slot, obtain $\mathcal{D}_{1}$:}\;
	$\mathcal{T}_{1}^{\rmct}=\emptyset$.\;
	$\mathcal{T}_{1}^{\rmtt}=\{(\mathbf{s}_{1|1}^{u},\mathbf{P}_{1|1}^{u})=\textrm{InitTrack}(\mathbf{z}_{1}^{u})|\forall  \mathbf{z}_{1}^{u}\in\mathcal{D}_{1}\}.$\;
	$\mathcal{T}_{1}=\mathcal{T}_{1}^{\rmct}\cup\mathcal{T}_{1}^{\rmtt}.$

	\ForEachTimeSlot{}{
		\ForEachTrackA{}{
			$(\mathbf{s}_{m|m-1}^{v},\mathbf{P}_{m|m-1}^{v})=\textrm{PredictTrack}(\tau_{m-1}^{v}).$\;
			Gating the detections by \eqref{eqn:gating}.\;
		}
		Solve Problem \textsf{P1} and derive $\mathcal{A}_{m}^{\rmct}$.\;
		$\widetilde{\mathcal{D}}_{m}^{\rmrem}=\mathcal{D}_{m}\backslash\mathcal{A}_{m}^{\rmct}$.\;
		Solve Problem \textsf{P2} and derive $\mathcal{A}_{m}^{\rmtt}$.\;
		$\mathcal{A}_{m}=\mathcal{A}_{m}^{\rmct}\cup\mathcal{A}_{m}^{\rmtt}$.\;
		$\mathcal{D}_{m}^{\rmrem}=\mathcal{D}_{m}\backslash\mathcal{A}_{m}$.\;
		Confirm and delete tracks based on the latest $N^{\rmcnf}$ and $N^{\rmdel}$ time slots, respectively, and derive the confirmed track subset  $\mathcal{T}_{m-1}^{\rmct,\prime}$ and tentative track subset $\mathcal{T}_{m-1}^{\rmtt,\prime}$.\;
		Initialize tracks as $\mathcal{T}_{m}^{\rminit}$ with $\mathcal{D}_{m}^{\rmrem}$ via \eqref{eqn:initTrackM}.\;
		\ForEachTrackB{}{
			\uIf{$\exists u, a_{m}^{u,v,\star}=1$}{
				$(\mathbf{s}_{m|m}^{v},\mathbf{P}_{m|m}^{v})=\textrm{CorrectTrack}(\tau_{m-1}^{v},\mathbf{z}_{m}^{u})$.\;
			}\uElse{
				$(\mathbf{s}_{m|m}^{v},\mathbf{P}_{m|m}^{v})=(\mathbf{s}_{m|m-1}^{v},\mathbf{P}_{m|m-1}^{v})$.\;
			}
		}
		Update the confirmed track set $\mathcal{T}_{m}^{\rmct}$ and tentative track set $\mathcal{T}_{m}^{\rmtt}$ by \eqref{eqn:updateCT} and \eqref{eqn:updateTT}, respectively.\;
		$\mathcal{T}_{m}=\mathcal{T}_{m}^{\rmct}\cup\mathcal{T}_{m}^{\rmtt}$.\;
		\Return{
			Set of CTs $\mathcal{T}_{m}^{\rmct}$ and set of TTs $\mathcal{T}_{m}^{\rmtt}$.\;
		}
	}
	\caption{The MTT framework}
	\label{alg:MTT}
\end{algorithm}

\subsection{Track Initialization}
\label{subsec:init}
When an observation is not associated with any track, i.e., the observations in the first time slot or the ones in $\mathcal{D}_{m}^{\rmrem}$ ($\forall m$), a new track will be initialized by the observation as denoted in \eqref{eqn:initTrack1} and \eqref{eqn:initTrackM}.

The track initialization for bistatic tracking is first elaborated.
According to the measurement model \eqref{eqn:measurementModel},
given the observation $\mathbf{z}_{m}^{u}$ in the $m$-th time slot, the state $\mathbf{s}_{m|m}^{u}$ and error covariance matrix $\mathbf{P}_{m|m}^{u}$ can be initialized as
\begin{subnumcases}{}
	\mathbf{s}_{m|m}^{u}&$=\mathbf{H}^{\sfT}\mathbf{z}_{m}^{u}$,                                                             \\
	\mathbf{P}_{m|m}^{u}&$=\diag(\sigma_{R}^{2},\sigma_{\dot{R}}^{2},\sigma_{0}^{2},\sigma_{\theta}^{2},\sigma_{0}^{2})$,
\end{subnumcases}
where $\mathbf{H}$ is defined in \eqref{eqn:H_bist}, $\sigma_{R}^{2}$, $\sigma_{\dot{R}}^{2}$, $\sigma_{\theta}^{2}$, and $\sigma_{0}^{2}$ are the variances of corresponding measurement errors.

On the other hand, when the Cartesian tracking is considered, the state of the new track $\mathbf{s}_{m|m}^{u}=[x_{m|m}^{u},\dot{x}_{m|m}^{u},y_{m|m}^{u},\dot{y}_{m|m}^{u}]^{\sfT}$, is initialized as
\begin{subequations}
	\begin{alignat}{2}
		\label{eqn:RrA2P}
		 & {}
		 &    & \big[x_{m|m}^{u},y_{m|m}^{u}\big]^{\sfT}\nonumber                                                             \\
		 & =
		 &    & \frac{(R_{m}^{u})^{2}-L_{\rmtxrx}^{2}}{2\big(R_{m}^{u}+L_{\rmtxrx}\cos(\theta_{m}^{u}-\theta_{\rmtxrx})\big)}
		\left[\begin{array}{c}
				      \!\cos\theta_{m}^{u}\! \\
				      \!\sin\theta_{m}^{u}\!
			      \end{array}\right]
		\!+\!
		\mathbf{l}_{\rmrx},
		\\
		\label{eqn:RrA2V}
		 & {}
		 &    & \big[\dot{x}_{m|m}^{u},\dot{y}_{m|m}^{u}\big]^{\sfT}\nonumber                                                 \\
		 & =
		 &    & \dot{R}_{m}^{u}\left.\frac{
			\frac{\mathbf{l}-\mathbf{l}_{\rmrx}}{\|\mathbf{l}-\mathbf{l}_{\rmrx}\|}
			+
			\frac{\mathbf{l}-\mathbf{l}_{\rmtx}}{\|\mathbf{l}-\mathbf{l}_{\rmtx}\|}
		}{
			\left\|
			\frac{\mathbf{l}-\mathbf{l}_{\rmrx}}{\|\mathbf{l}-\mathbf{l}_{\rmrx}\|}
			+
			\frac{\mathbf{l}-\mathbf{l}_{\rmtx}}{\|\mathbf{l}-\mathbf{l}_{\rmtx}\|}
			\right\|
		}\right|_{\begin{subarray}{l}
			          \mathbf{l}=\big[x_{m|m}^{u},y_{m|m}^{u}\big]^{\sfT}
		          \end{subarray}}.
	\end{alignat}
\end{subequations}
With the assumption that the measurement errors of bistatic range, bistatic range rate and AoA are independent, i.e., $\mathbf{R}=\diag(\sigma_{R}^{2},\sigma_{\dot{R}}^{2},\sigma_{\theta}^{2})$ in \eqref{eqn:measurementModel}, the covariance matrix is initialized as
\begin{equation}
	\mathbf{P}_{m|m}^{v}
	=
	\left[
		\begin{array}{cccc}
			\sigma_{x}^{2} & 0                    & \sigma_{xy}    & 0                    \\
			0              & \sigma_{\dot{x}}^{2} & 0              & 0                    \\
			\sigma_{xy}    & 0                    & \sigma_{y}^{2} & 0                    \\
			0              & 0                    & 0              & \sigma_{\dot{y}}^{2} \\
		\end{array}
		\right],
\end{equation}
where $\sigma_{x}^{2}$ and $\sigma_{y}^{2}$ denote the variances of estimation error in the x-coordinate and y-coordinate of the target respectively, $\sigma_{xy}$ is their covariance, $\sigma_{\dot{x}}^{2}$ and $\sigma_{\dot{y}}^{2}$ denotes the variances of velocity estimation errors.
Moreover,
\begin{subequations}
	\begin{alignat}{3}
		 & \sigma_{x}^{2} &  & = &  & \frac{H_{1}^{2}\sigma_{R}^{2}+H_{2}^{2}\sigma_{\theta}^{2}}{H_{3}^{4}},   \\
		 & \sigma_{y}^{2} &  & = &  & \frac{H_{4}^{2}\sigma_{R}^{2}+H_{5}^{2}\sigma_{\theta}^{2}}{H_{3}^{4}},   \\
		 & \sigma_{xy}    &  & = &  & \frac{H_{1}H_{4}\sigma_{R}^{2}+H_{2}H_{5}\sigma_{\theta}^{2}}{H_{3}^{4}},
	\end{alignat}
\end{subequations}
where
\begin{subequations}
	\begin{alignat}{3}
		 & H_{1} &  & = &  & 2RL_{\rmtxrx}\cos^{2}\theta+(R^{2}+L_{\rmtxrx}^{2})\cos\theta|_{\mathbf{z}=\mathbf{z}_{m}^{u}}, \\
		 & H_{2} &  & = &  & (L_{\rmtxrx}^{2}-R^{2})R\sin\theta|_{\mathbf{z}=\mathbf{z}_{m}^{u}},                            \\
		 & H_{3} &  & = &  & \sqrt{2}(L_{\rmtxrx}\cos\theta+R)|_{\mathbf{z}=\mathbf{z}_{m}^{u}},                             \\
		 & H_{4} &  & = &  & RL_{\rmtxrx}\sin2\theta+(R^{2}+L_{\rmtxrx}^{2})\sin\theta|_{\mathbf{z}=\mathbf{z}_{m}^{u}},     \\
		 & H_{5} &  & = &  & (R^{2}-L_{\rmtxrx}^{2})(R\cos\theta+L_{\rmtxrx})|_{\mathbf{z}=\mathbf{z}_{m}^{u}}.
	\end{alignat}
\end{subequations}

\subsection{Track Prediction}
\label{subsec:trackPrediction}
The track prediction in \eqref{eqn:predictTrack} provides prior estimation of the target's state in the $m$-th time slot, providing a track in the $(m-1)$-th time slot \cite{bishop2001introduction}.
First, the state transition model for both tracking methods can be written as
\begin{equation}
	\label{eqn:transitionModel}
	\mathbf{s}_{m}=\mathbf{F}\mathbf{s}_{m-1}+\mathbf{w}_{m},
\end{equation}
where $\mathbf{F}$ is the state transition matrix, and $\mathbf{w}_{m}\sim\mathcal{N}(\mathbf{0},\mathbf{Q})$ is the noise with covariance matrix $\mathbf{Q}$.
Hence, in the $m$-th time slot, the prediction of state $\mathbf{s}_{m|m-1}^{v}$ and corresponding error covariance matrix $\mathbf{P}_{m|m-1}^{v}$ in \eqref{eqn:predictTrack} can be obtained as
\begin{subnumcases}{}
	\mathbf{s}_{m|m-1}^{v}&$=\mathbf{F}\mathbf{s}_{m-1|m-1}^{v}$,\\
	\mathbf{P}_{m|m-1}^{v}&$=\mathbf{F}\mathbf{P}_{m-1|m-1}^{v}\mathbf{F}^{\sfT}+\mathbf{Q}$,
\end{subnumcases}
where $(\mathbf{s}_{m-1|m-1}^{v},\mathbf{P}_{m-1|m-1}^{v}) \in \tau_{m-1}^{v}$.

When the bistatic tracking is considered, the state transition matrix  $\mathbf{F}$ is represented by
\begin{equation}
	\mathbf{F}
	\triangleq
	\left[\begin{array}{ccccc}
			1 & T & T^{2}/2 & 0 & 0 \\
			0 & 1 & T       & 0 & 0 \\
			0 & 0 & 1       & 0 & 0 \\
			0 & 0 & 0       & 1 & T \\
			0 & 0 & 0       & 0 & 1 \\
		\end{array}\right].
\end{equation}
Moreover, the noise covariance $\mathbf{Q}=\diag\big(\sigma_{\ddot{R}}^{2}\mathbf{Q}_{R},\sigma_{\ddot{\theta}}^{2}\mathbf{Q}_{\theta}\big)$, where $\sigma_{\ddot{R}}^{2}$ and $\sigma_{\ddot{\theta}}^{2}$ are the variances of estimation errors for range acceleration and AoA acceleration, and
\begin{equation}
	\mathbf{Q}_{R}
	=
	\left[
		\begin{array}{ccc}
			T^{4}/4 & T^{3}/2 & T^{2}/2 \\
			T^{3}/2 & T^{2}   & T       \\
			T^{2}/2 & T       & 1       \\
		\end{array}
		\right],
\end{equation}
\begin{equation}
	\mathbf{Q}_{\theta}
	=
	\left[
		\begin{array}{cc}
			T^{4}/4 & T^{3}/2 \\
			T^{3}/2 & T^{2}   \\
		\end{array}
		\right].
\end{equation}

On the other hand, when the Cartesian tracking is considered, the state transition matrix is given by
\begin{equation}
	\mathbf{F}
	\triangleq
	\left[\begin{array}{cccc}
			1 & T & 0 & 0 \\
			0 & 1 & 0 & 0 \\
			0 & 0 & 1 & T \\
			0 & 0 & 0 & 1 \\
		\end{array}\right],
\end{equation}
and the noise covariance $\mathbf{Q}=\diag\big(\sigma_{\ddot{x}}^{2}\mathbf{Q}_{x},\sigma_{\ddot{y}}^{2}\mathbf{Q}_{y}\big)$, where $\sigma_{\ddot{x}}^{2}$ and $\sigma_{\ddot{y}}^{2}$ are the variances of estimation errors for x-axis and y-axis accelerations respectively,
\begin{equation}
	\mathbf{Q}_{x}
	=
	\mathbf{Q}_{y}
	=
	\left[
		\begin{array}{cc}
			T^{4}/4 & T^{3}/2 \\
			T^{3}/2 & T^{2}   \\
		\end{array}
		\right].
\end{equation}

\subsection{Track Correction}
\label{subsec:trackCorrection}
The track correction operator in \eqref{eqn:updateTrackA} provides a refined state of the target in the $m$-th time slot, given a track in the $(m-1)$-th time slot and an associated observation in the $m$-th time slot \cite{bishop2001introduction}.

The track correction for the bistatic tracking is first elaborated.
With the linear measurement model in \eqref{eqn:measurementModel}, the refined state $\mathbf{s}_{m|m}^{v}$ and corresponding covariance matrix $\mathbf{P}_{m|m}^{v}$ of a track $\tau_{m-1}^{v}$ and an associated observation $\mathbf{z}_{m}^{u}$ ($a_{m}^{u,v,\star}=1$) in \eqref{eqn:updateTrackA} can be calculated by
\begin{subnumcases}{}
	\mathbf{s}_{m|m}^{v}
	=\mathbf{s}_{m|m-1}^{v}+\mathbf{K}_{m}^{v}(\mathbf{z}_{m}^{u}-\mathbf{H}\mathbf{s}_{m|m-1}^{v}),\\
	\mathbf{P}_{m|m}^{v}
	=(\mathbf{I}-\mathbf{K}_{m}^{v}\mathbf{H})\mathbf{P}_{m|m-1}^{v},
\end{subnumcases}
where the optimal Kalman gain $\mathbf{K}_{m}^{v}$ is given by
\begin{align}
	\mathbf{K}_{m}^{v}=\mathbf{P}_{m|m-1}^{v}\mathbf{H}^{\sfT}[\mathbf{W}_{m}^{v}]^{-1}.
\end{align}

For the Cartesian tracking, with the measurement model in \eqref{eqn:measurementModel} and \eqref{eqn:measurementFunctionCartesian}, the refined state $\mathbf{s}_{m|m}^{v}$ and corresponding covariance matrix $\mathbf{P}_{m|m}^{v}$ of a track $\tau_{m-1}^{v}$ and an associated observation $\mathbf{z}_{m}^{u}$ ($a_{m}^{u,v,\star}=1$) in \eqref{eqn:updateTrackA} can be calculated by
\begin{subnumcases}{}
	\mathbf{s}_{m|m}^{v}=\mathbf{s}_{m|m-1}^{v}+\mathbf{K}_{m}^{v}(\mathbf{z}_{m}^{u}-\mathbf{h}(\mathbf{s}_{m|m-1}^{v}))\\
	\mathbf{P}_{m|m}^{v}=(\mathbf{I}-\mathbf{K}_{m}^{v}\mathbf{H}_{m}^{v})\mathbf{P}_{m|m-1}^{v},
\end{subnumcases}
where the near-optimal Kalman gain is written by
\begin{equation}
	\mathbf{K}_{m}^{v}=\mathbf{P}_{m|m-1}^{v}[\mathbf{H}_{m}^{v}]^{\sfT}[\mathbf{W}_{m}^{v}]^{-1},
\end{equation}
the Jacobian matrix of the measurement function in the $m$-th time slot is given by
\begin{align}
	\mathbf{H}_{m}^{v}
	=
	\left.\left[
		\begin{array}{cccccc}
			\frac{\partial R}{\partial x}       & 0                                         &
			\frac{\partial R}{\partial y}       & 0
			\\
			\frac{\partial \dot{R}}{\partial x} & \frac{\partial \dot{R}}{\partial \dot{x}} &
			\frac{\partial \dot{R}}{\partial y} & \frac{\partial \dot{R}}{\partial \dot{y}}
			\\
			\frac{\partial \theta}{\partial x}  & 0                                         &
			\frac{\partial \theta}{\partial y}  & 0
			\\
		\end{array}
		\right]\right|_{\mathbf{s}=\mathbf{s}_{m|m-1}^{v}},
\end{align}
and the partial derivatives are given by
\begin{subequations}
	\begin{alignat}{3}
		 &
		\frac{\partial R}{\partial x}
		 &   &
		=
		 &   &
		\frac{\partial \dot{R}}{\partial \dot{x}}
		=
		\frac{x-x_{\rmrx}}{\|\mathbf{l}-\mathbf{l}_{\rmrx}\|}
		+
		\frac{x-x_{\rmtx}}{\|\mathbf{l}-\mathbf{l}_{\rmtx}\|},
		\\
		 &
		\frac{\partial \dot{R}}{\partial x}
		 &   &
		=
		 &   &
		\frac{\dot{x}}{\|\mathbf{l}-\mathbf{l}_{\rmrx}\|}
		-
		\frac{(x-x_{\rmrx})(\mathbf{l}-\mathbf{l}_{r})^{\sfT}\mathbf{v}}{\|\mathbf{l}-\mathbf{l}_{\rmrx}\|^{3}}
		\nonumber
		\\
		 &
		 &   &
		 &   &
		+
		\frac{\dot{x}}{\|\mathbf{l}-\mathbf{l}_{\rmtx}\|}
		-
		\frac{(x-x_{\rmtx})(\mathbf{l}-\mathbf{l}_{t})^{\sfT}\mathbf{v}}{\|\mathbf{l}-\mathbf{l}_{\rmtx}\|^{3}},
		\\
		 &
		\frac{\partial \theta}{\partial x}
		 &   &
		=
		 &   &
		-\frac{y-y_{\rmrx}}{\|\mathbf{l}-\mathbf{l}_{\rmrx}\|^{2}}.
	\end{alignat}
\end{subequations}
Moreover, $\frac{\partial R}{\partial y}\!=\!\frac{\partial \dot{R}}{\partial \dot{y}}$, $\frac{\partial \dot{R}}{\partial y}$ and $\frac{\partial \theta}{\partial y}$ can be expressed similarly.

\section{Experimental Results and Discussions}
\label{sec:experiment}
\subsection{System Configuration}
\begin{figure}[htb]
	\centering
	\includegraphics[width=0.8\linewidth]{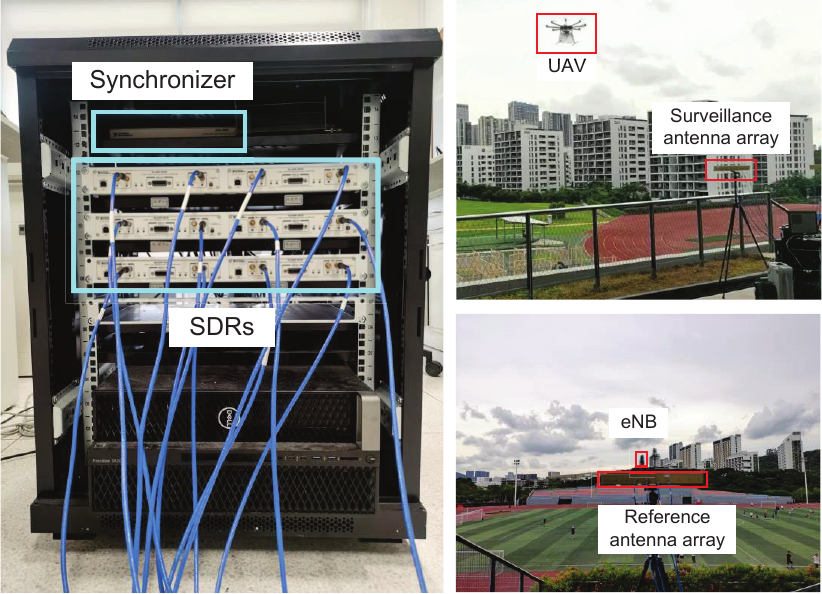}
	\caption{The experimental platform of LIPASE.}
	\label{fig:hardware}
\end{figure}

The experiment scenario and the LIPASE receiver are illustrated in Fig. \ref{fig:hardware}, where two ULAs with half-wavelength inter-element spacing are connected with six software-defined radio (SDR) as the reference and surveillance antenna arrays, respectively.
The reference antenna array consists of four antenna elements, while the surveillance antenna array consists of eight antenna elements.
There are more antenna elements in the latter for better AoA detection resolution.
The received signals of all twelve antenna elements are sampled by the SDRs with synchronized timing provided by a synchronizer.
All the sampled baseband signals are then saved in a server for data processing.

In the experiments, the LTE-FDD downlink at the 2130--2135 MHz frequency band is chosen for passive sensing.
The parameters of system configuration and the proposed tracking framework are summarized in Table \ref{tab:parameter}.

\begin{table*}[!htbp]
	\small
	\centering
	\caption{Experimental parameters.}
	\label{tab:parameter}
	\begin{tabular}{|c|c|c|c|}
		\hline
		\multicolumn{2}{|c|}{\textbf{Parameter}} & \textbf{Symbol}                                       & \textbf{Value}                                                                                                                                   \\
		\hline
		\multirow{8}{*}{System}
		{}                                       & Baseline length                                       & $L_{\rmtxrx}$                                             & 255 m                                                                                \\
		\cline{2-4}
		{}                                       & Reference array size                                  & $N_{\rmref}$                                              & 4                                                                                    \\
		\cline{2-4}
		{}                                       & Surveillance array size                               & $N_{\rmsur}$                                              & 8                                                                                    \\
		\cline{2-4}
		{}                                       & Carrier frequency                                     & $f_{\mathrm{c}}$                                          & 2123 MHz                                                                             \\
		\cline{2-4}
		{}                                       & Sampling rate                                         & $f_{\mathrm{s}}$                                          & 25 MHz                                                                               \\
		\cline{2-4}
		{}                                       & Operation frequency                                   & $/$                                                       & 2130-2135 MHz                                                                        \\
		\cline{2-4}
		{}                                       & Bandwidth                                             & $B$                                                       & 5 MHz                                                                                \\
		\cline{2-4}
		{}                                       & CIT                                                   & $T$                                                       & 0.2 s                                                                                \\
		\hline
		\multirow{2}{*}{CC}
		{}                                       & Maximal Doppler shift                                 & $P$                                                       & 0                                                                                    \\
		\cline{2-4}
		{}                                       & Maximal sample delay                                  & $L$                                                       & 20                                                                                   \\
		\hline
		\multirow{4}{*}{CA-CFAR}
		{}                                       & \tabincell{c}{\# of guard cells                                                                                                                                                                          \\
		(\# of Doppler bins, \# of range bins)}  & $(P_{\rmguard},L_{\rmguard})$                         & $(60,1)$                                                                                                                                         \\
		\cline{2-4}
		{}                                       & \tabincell{c}{\# of training cells                                                                                                                                                                       \\
		(\# of Doppler bins, \# of range bins)}  & $(P_{\rmtrain},L_{\rmtrain})$                         & $(60,1)$                                                                                                                                         \\
		\cline{2-4}
		{}                                       & Threshold factor                                      & $\alpha$                                                  & 15 dB                                                                                \\
		\hline
		\multirow{6}{*}{MTT}
		{}                                       & \# of time slot w/ association for track confirmation & $N^{\rmcnf}$                                              & $5$                                                                                  \\
		\cline{2-4}
		{}                                       & \# of time slot w/o association for track deletion    & $N^{\rmdel}$                                              & $14$                                                                                 \\
		\cline{2-4}
		{}                                       & Gating threshold                                      & $\gamma$                                                  & 20                                                                                   \\
		\cline{2-4}
		{}                                       & Process noise covariance for bistatic tracking        & $\sigma_{\ddot{R}}^{2}$, $\sigma_{\ddot{\theta}}^{2}$     & $10^{2}\ \mathrm{m}^{2}/\mathrm{s}^{2}$, $3^{2}\ \mathrm{deg}^{2}$                   \\
		\cline{2-4}
		{}                                       & Process noise covariance for Cartesian tracking       & $\sigma_{\ddot{x}}^{2}$, $\sigma_{\ddot{y}}^{2}$          & $4^{2}\ \mathrm{m}^{2}/\mathrm{s}^{4}$, $4^{2}\ \mathrm{m}^{2}/\mathrm{s}^{4}$       \\
		\cline{2-4}
		{}                                       & Measurement noise covariance                          & $\sigma_{R}^{2},\sigma_{\dot{R}}^{2},\sigma_{\theta}^{2}$ & $7^{2}\ \mathrm{m}^{2},1^{2}\ \mathrm{m}^{2}/\mathrm{s}^{2},3^{2}\ \mathrm{deg}^{2}$ \\
		\hline
	\end{tabular}
\end{table*}

\begin{figure}[htb]
	\centering
	\includegraphics[width=0.78\linewidth]{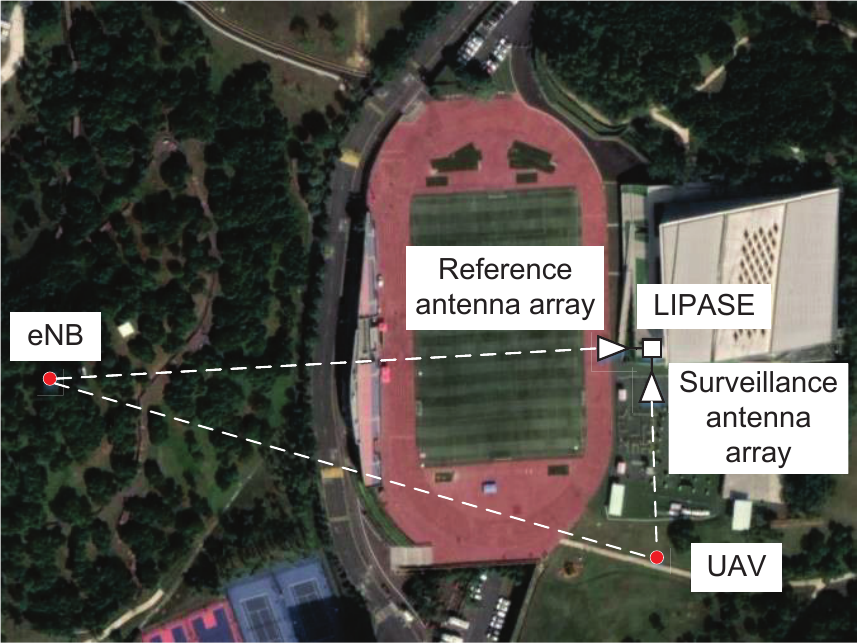}
	\caption{Overview of experiment scenario.}
	\label{fig:system_deployment}
\end{figure}

The overview of the experiment scenario is shown in Fig. \ref{fig:system_deployment}.
LIPASE is deployed on the platform of a building which is $255$ meters away from the LTE eNB located on a hill.
We control a quadcopter hung with a steel sphere to imitate a delivery drone with a payload.
The flight trajectory follows a `J' shape.
Due to the low localization accuracy of the intrinsic global positioning system (GPS) system, which is not sufficiently accurate as the experiment ground truth, the drone is equipped with an additional differential GPS (DGPS) module.
The DGPS reference station is deployed on the ground, such that sub-meter localization accuracy can be achieved.

\subsection{Range, Velocity and Angular Resolutions}
The range resolution $\Delta R$ is the minimum required range to distinguish two different targets, which is defined by \cite{griffiths2022introduction}
\begin{equation}
	\Delta R = \frac{c}{2B\cos(\beta^{\rmbist}/2)},
\end{equation}
where $B$ is the bandwidth and $\beta^{\rmbist}$ is the bistatic angle.
In our experimental configuration with a bandwidth of $5$ MHz, the range resolution is $30$ m at $\beta^{\rmbist}=0$°.

The Doppler resolution $\Delta f_{\rmD}$ indicates the capability of a radar system to differentiate two targets with different Doppler frequencies, which depends on the CIT \cite{griffiths2022introduction}.
Particularly,
\begin{equation}
	\Delta f_{\rmD}=\frac{1}{T}=\frac{f_{\mathrm{s}}}{N}.
\end{equation}
Then the velocity resolution is given by
\begin{equation}
	\Delta v=\frac{\lambda\Delta f_{\rmD}}{2\cos(\beta^{\rmbist}/2)}=\frac{\lambda}{2T\cos(\beta^{\rmbist}/2)}.
\end{equation}
In our experiment, the CIT is $T=0.2$ s, the Doppler resolution is $5$ Hz, and the velocity resolution of $0.35$ m/s at $\beta^{\rmbist}=0$°.

The angular resolution of an antenna array, denoted by $\Delta \theta$, indicates the capability of the antenna array to differentiate two targets in the space domain.
The angular resolution of a ULA with isotropic elements and half-wavelength element spacing can be approximately estimated by \cite{tse2005fundamentals}
\begin{equation}
	\Delta \theta=\frac{2}{N|\cos(\theta)|},
\end{equation}
where $\theta$ denotes the angle between the target direction and the broadside of the ULA.
The 8-element surveillance antenna array is with an angular resolution of $14.3$° at the broadside.
In the following parts, it is shown that with the MTT algorithm, the accuracy of trajectory tracking is much better than the above resolutions.

\subsection{Detection and Tracking Results}
\label{subsec:trackingRDA}
\begin{figure*}[htb]
	\centering
	\subfloat[]{\includegraphics[width=0.33\linewidth]{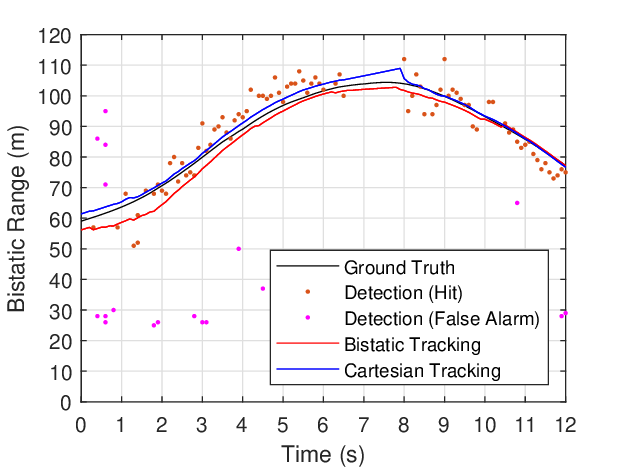}}
	\subfloat[]{\includegraphics[width=0.33\linewidth]{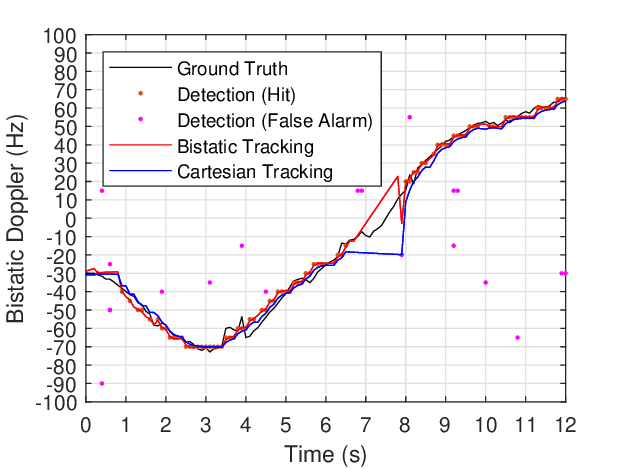}}
	\subfloat[]{\includegraphics[width=0.33\linewidth]{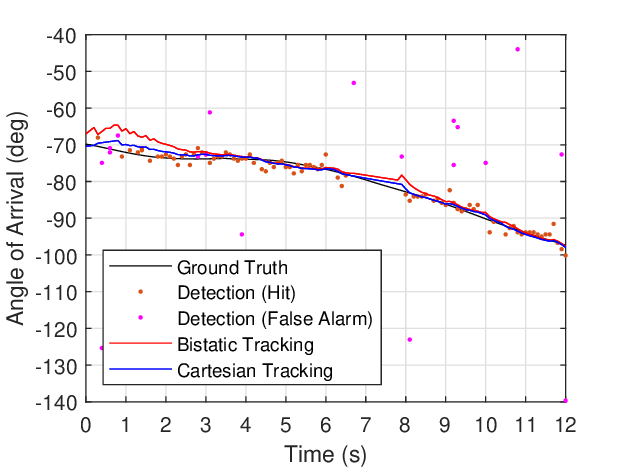}}
	\caption{Observations and tracking results versus time in terms of
		(a) The bistatic range, (b) bistatic Doppler shift, and (c) AoA.}
	\label{fig:time-RDA}
\end{figure*}
\begin{figure}[htb]
	\centering
	\subfloat[]{\includegraphics[width=0.8\linewidth]{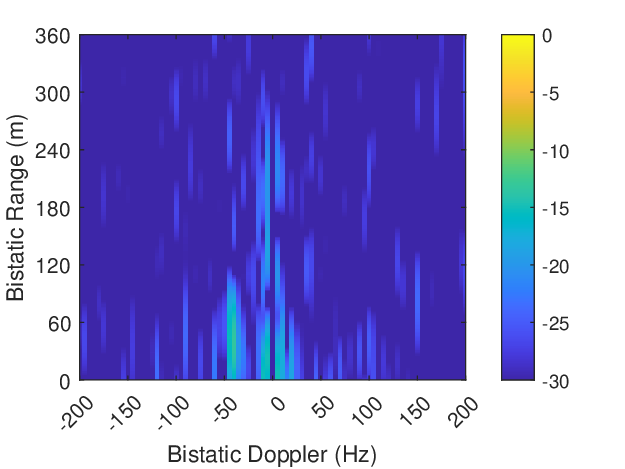}}\\
	\subfloat[]{\includegraphics[width=0.8\linewidth]{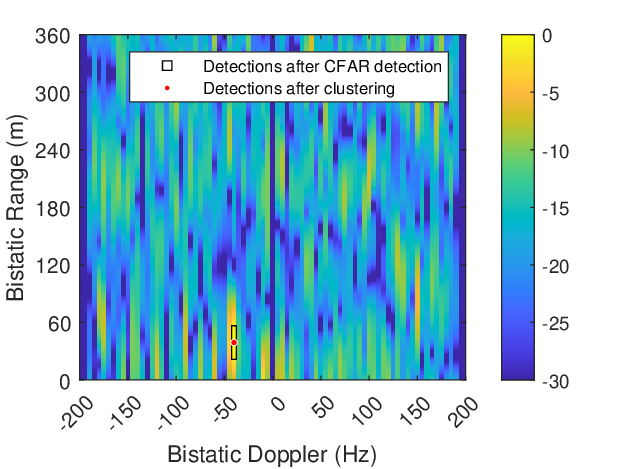}}
	\caption{A sample of the RD response (a) before and (b) after CFAR detection and clustering.}
	\label{fig:CFAR}
\end{figure}

A sample of RD response of one raw surveillance signal in a time slot is illustrated in Fig. \ref{fig:CFAR}, together with the response after CFAR detection.
It can be observed that after CFAR detection, targets with weak echo signal strength can still be detected.

The detection and tracking results in terms of the bistatic range, Doppler frequency and AoA are illustrated in Fig. \ref{fig:time-RDA}, where the observations versus time slot, the ground truth trajectory, and the estimated trajectories via both methods are plotted and compared.
It can be observed that due to the limited resolutions in the measurements of bistatic range, Doppler frequency and AoA, the detection rate of the target UAV is only 71.9\%.
Moreover, the overall missed detection rate is 28.1\%, and the false alarm rate is 18.2\% during the measurement period.
The presence of missed detections during the time interval of 6.5--8.0 s is because the UAV's trajectory follows the bistatic contour (i.e., zero bistatic Doppler shift).

To evaluate the measurement and tracking accuracy, we use the mean absolute error (MAE) and root mean squared error (RMSE) as the metrics.
The MAEs and RMSE of an estimated value $\hat{x}_{m}$ deviated from the ground truth value $x_{m}$ in the $M$ time slots are evaluated by
\begin{alignat}{3}
	 & \textrm{MAE}  &  & = &  & \frac{1}{M}\sum_{m=1}^{M}|\hat{x}_{m}-x_{m}|,            \\
	 & \textrm{RMSE} &  & = &  & \sqrt{\frac{1}{M}\sum_{m=1}^{M}(\hat{x}_{m}-x_{m})^{2}}.
\end{alignat}
The MAEs and RMSEs of detections and tracking results are listed in Table \ref{table:errorRDA}.

\begin{table}[tb]
	\small
	\centering
	\resizebox{\linewidth}{!}{%
		\begin{tabular}{|c|c|c|c|c|c|c|}
			\hline
			\multirow{2}{*}{\scriptsize{\diagbox[width=80pt,height=24pt]{Method}{Value}{Error}}} 
			                   & \multicolumn{2}{c|}{Range (m)}
			                   & \multicolumn{2}{c|}{Doppler (Hz)}
			                   & \multicolumn{2}{c|}{AoA (°)}                                                                                      \\
			\cline{2-7}
			{}                 & MAE                               & RMSE          & MAE           & RMSE          & MAE           & RMSE          \\
			\hline
			Detection (Hit)    & 4.40                              & 5.37          & \textbf{2.59} & \textbf{3.26} & 1.14          & 1.55          \\
			\hline
			Bistatic Tracking  & 2.43                              & 2.97          & 3.27          & 4.67          & 1.76          & 2.45          \\
			\hline
			Cartesian Tracking & \textbf{1.44}                     & \textbf{1.78} & 4.21          & 6.91          & \textbf{0.80} & \textbf{1.02} \\
			\hline
		\end{tabular}
	}
	\caption{MAEs and RMSEs of bistatic range, Doppler frequency and AoA.}
	\label{table:errorRDA}
\end{table}

\begin{figure*}[t!]
	\centering
	\begin{minipage}{0.99\textwidth}
		\centering
		\subfloat[]{\includegraphics[width=0.24\linewidth]{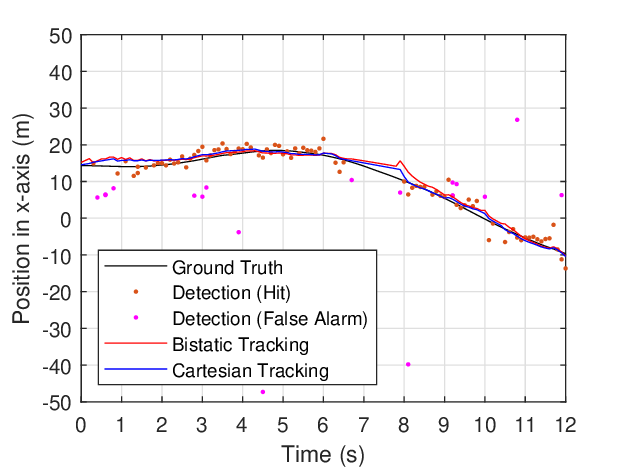}}
		\subfloat[]{\includegraphics[width=0.24\linewidth]{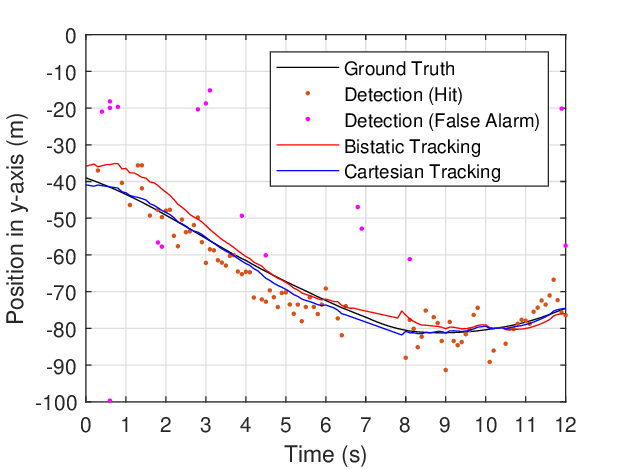}}
		\subfloat[]{\includegraphics[width=0.24\linewidth]{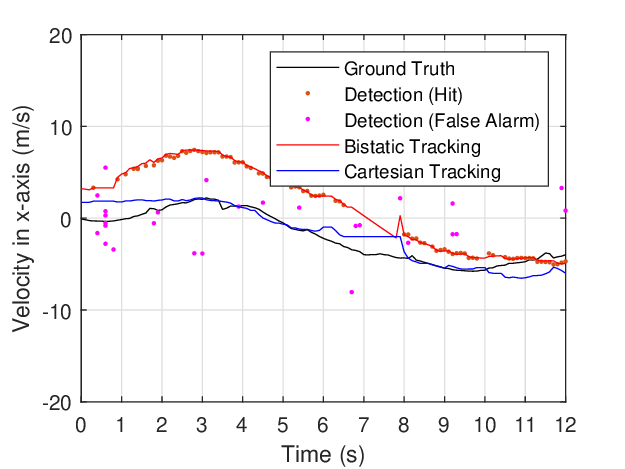}}
		\subfloat[]{\includegraphics[width=0.24\linewidth]{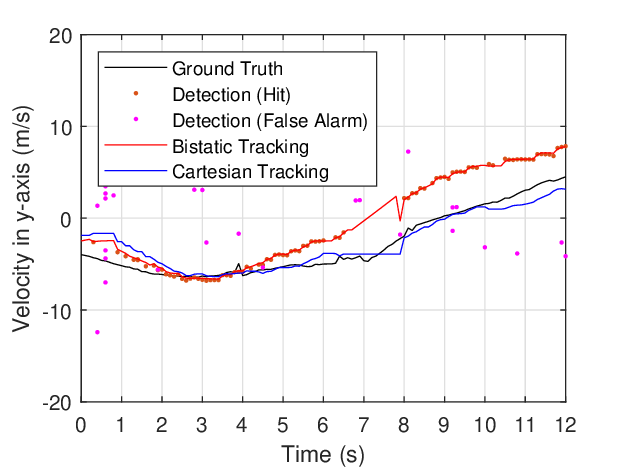}}
		\caption{Observations and tracking results versus time in terms of (a) x-axis position, (b) y-axis position, (c) x-axis velocity and (d) y-axis velocity.}
		\label{fig:time-PV}
	\end{minipage}
	\hfill
	\footnotesize
	\centering
	\begin{tabular}{|c|c|c|c|c|c|c|c|c|c|c|}
		\hline
		\multirow{3}{*}{\diagbox[width=80pt,height=36pt]{Method}{Value}{Error}} 
		                   & \multicolumn{4}{c|}{Position (m)}
		                   & \multicolumn{4}{c|}{Velocity (m/s)}
		                   & \multicolumn{2}{c|}{\multirow{2}{*}{Localization (m)}}                                                                                                                                                 \\
		\cline{2-9}
		{}
		                   & \multicolumn{2}{c|}{X-axis}
		                   & \multicolumn{2}{c|}{Y-axis}
		                   & \multicolumn{2}{c|}{X-axis}
		                   & \multicolumn{2}{c|}{Y-axis}
		                   & \multicolumn{2}{c|}{}                                                                                                                                                                                  \\
		\cline{2-11}
		{}                 & MAE                                                    & RMSE          & MAE           & RMSE          & MAE           & RMSE          & MAE           & RMSE          & MAE           & RMSE          \\
		\hline
		Detection (Hit)    & 1.40                                                   & 1.92          & 3.73          & 4.56          & 3.37          & 3.83          & 2.16          & 2.66          & 4.18          & 4.94          \\
		\hline
		Bistatic Tracking  & 1.17                                                   & 1.43          & 2.47          & 3.23          & 3.43          & 3.85          & 2.44          & 2.88          & 2.85          & 3.53          \\
		\hline
		Cartesian Tracking & \textbf{0.84}                                          & \textbf{1.02} & \textbf{0.84} & \textbf{1.09} & \textbf{0.97} & \textbf{1.24} & \textbf{0.87} & \textbf{1.19} & \textbf{1.33} & \textbf{1.49} \\
		\hline
	\end{tabular}
	\caption{MAEs and RMSEs of position and velocity in Cartesian plane.}
	\label{table:errorPV}
\end{figure*}

The detection and tracking results in terms of the position and velocity of both the x and y axes are illustrated in Fig. \ref{fig:time-PV}, while the corresponding MAEs and RMSEs are compared in Table \ref{table:errorPV}.
It can be observed that although bistatic tracking can achieve the lower MAE and RMSE of Doppler frequency in Table \ref{table:errorRDA}, it has a larger estimation error on the velocity and position than Cartesian tracking.
Hence, the proposed Cartesian tracking reaches the best accuracy in both velocity estimation and localization.
This is because the state transition model of the Cartesian tracking might be closer to the real experiment.

\begin{figure}[H]
	\centering
	\includegraphics[width=0.7\linewidth]{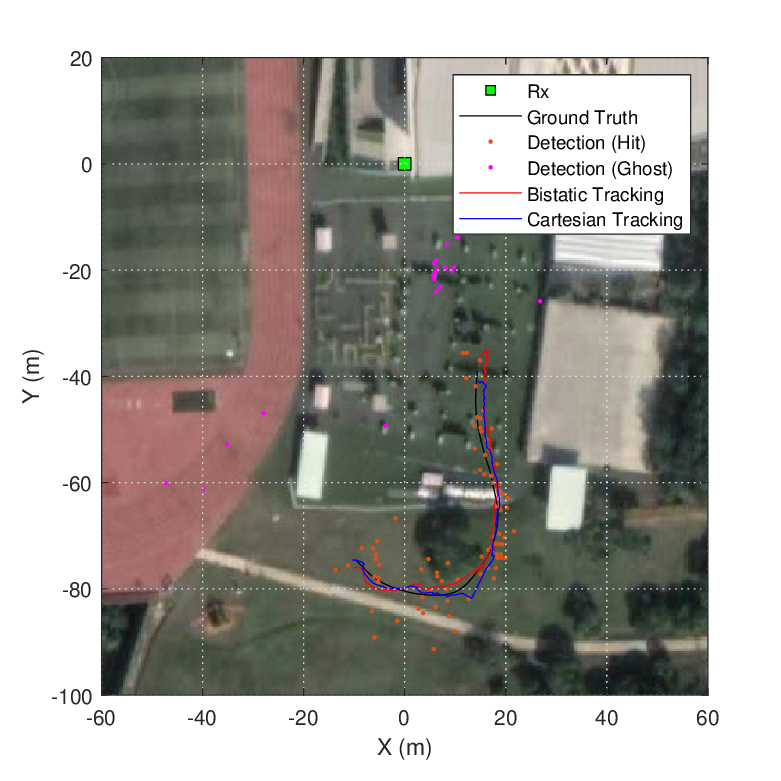}
	\caption{The estimated trajectories of the drone.}
	\label{fig:trackingCartesian}
\end{figure}

Finally, the estimated trajectories of LIPASE (by two tracking methods) and the ground truth are all shown in Fig. \ref{fig:trackingCartesian}, which shows that the proposed tracking algorithm can not only address the presence of missed detections and false alarms but also smooth the distorted detections.

\section{Conclusion}
\label{sec:conclusion}
In this paper, we propose an LTE-based passive sensing system with digital arrays, namely LIPASE, to track the trajectory of a flying UAV.
Because of the significant issue of false alarms and missed detections, a multi-target tracking algorithm is proposed to reconstruct the trajectory with the observations of passive sensing.
Both bistatic and Cartesian tracking methods are used in the multi-target tracking algorithm.
It is shown by experiments that LIPASE can achieve a meter-level localization error with Cartesian tracking, demonstrating that LIPASE could fulfill the demands of UAV flight tracking.

\bibliographystyle{IEEEtran}
\bibliography{IEEEabrv,reference}
\end{document}